\documentclass{emulateapj}

\usepackage{amsmath, amsthm, amssymb,epsfig,rotating}
\usepackage{natbib}
\usepackage{graphicx}
\usepackage{epstopdf}
\usepackage{enumerate}

\newcommand{\etal}{{\it et al.}}
\newcommand{\eg}{{\it e.g.,}}
\newcommand{\ie}{{\it i.e.,}}
\newcommand{\chisq}{{$\chi^{2}$}}
\newcommand{\msig}{{M$_{\rm BH} - \sigma_{*}$}}
\newcommand{\mlum}{{M$_{\rm BH} - $L$_{\rm bulge}$}}

\newcommand{\ssig}{{$\sigma_{*}$}}

\begin{document}

\title{Kinematics of Stellar Populations in Post-Starburst Galaxies}

\author{Kyle D. Hiner\altaffilmark{1,2}, Gabriela Canalizo\altaffilmark{1}}

\altaffiltext{1}{Department of Physics and Astronomy, University of California, Riverside, CA 92521, USA; email: gabriela.canalizo@ucr.edu}
\altaffiltext{2}{Departamento de Astronom\'{i}a, Universidad de Concepci\'{o}n, Chile; email: khiner@astro-udec.cl}

\begin{abstract}

Post-starburst galaxies host a population of early-type stars (A or F), but simultaneously lack indicators of ongoing star formation such as [\ion{O}{2}] emission. Two distinct stellar populations have been identified in these systems: a young post-starburst population superimposed on an older host population. We present a study of nine post-starburst galaxies with the following objectives: 1) to investigate if and how kinematical differences between the young and old populations of stars can be measured; and 2) to gain insight into the formation mechanism of the young population in these systems. We fit high signal-to-noise spectra with two independent populations in distinct spectral regions: the Balmer region, the \ion{Mg}{1b} region, and the Ca Triplet when available. We show that the kinematics of the two populations largely track one another if measured in the Balmer region with high signal-to-noise data. Results from examining the Faber-Jackson relation and the Fundamental Plane indicate these objects are not kinematically disturbed relative to more evolved spheroids. A case-study of the internal kinematics of one object in our sample shows it to be pressure-supported and not rotationally-dominated. Overall our results are consistent with merger-induced starburst scenarios where the young population is observed during the later stages of the merger. 

\end{abstract}


\singlespace

\section{Introduction}

Post-starburst galaxies (also known as ``E+A'' or ``K+A'' galaxies) have been an intriguing class of objects since their discovery by \citet{DresslerGunn83}. The class is defined as having elliptical or S0 morphologies and spectra that exhibit A or F type stars, but the spectra also lack ongoing star formation indicators (\eg~[\ion{O}{2}] $\lambda$3727, H$\alpha$ emission). The physical interpretation of these systems that has emerged invokes a scenario where two or more progenitor galaxies have interacted or merged. This merger triggers a large burst of star formation which was extinguished within the last $ < 1$ Gyr. Galaxy mergers have also been invoked as a mechanism that may contribute to establishing the black hole - host galaxy relations (\ie~\msig, \mlum) regardless of whether the shutdown of star formation was caused by feedback from an active nucleus \citep[see][]{Peng07,Jahnke11,CanoDiaz12, AH12}. Thus, the study of post-starburst galaxies may provide additional clues to galaxy evolution, and it is important to determine their relationship, if any, to both active and quiescent galaxies. 

Evidence for the interaction-driven origin of post-starburst galaxies has grown steadily over the years. However, whether those interactions are major mergers or simply close encounters is still a matter of debate. \citet{Zabludoff96} find tidal features in the morphologies of post-starburst galaxies in the field consistent with previous and more current galaxy merger simulations \citep[\eg][]{Ji14,Mihos94}. Under a merger scenario, dissipative star-forming gas is driven to the center of the potential. Therefore, one might expect the starburst formed during the merger to be centrally concentrated within the host galaxy. Several studies have attempted to measure the concentration of the post-starburst population relative to the underlying host with mixed results. This is often attempted by measuring Balmer absorption strengths as a function of radius (Balmer gradients). As noted by \citet{Pracy13}, the ability to compare the populations depends on the intrinsic spatial distribution of the post-starburst population, the observed image quality (seeing), and the spatial distribution of the continuum light. 

\citet{Snyder11} study merging galaxy models to reconcile the observed incidence of post-starburst galaxies with merger rates. They find that Balmer absorption lines are strongest in the central kiloparsec of the merger remnant, indicating that the post-starburst population is nuclear in origin. Interestingly, Snyder \etal~find that while the post-starburst lifetime is sensitive to many of the initial parameters of the merger models, for a major merger of equal-mass, gas-rich progenitors, the observability of the post-starburst phase is typically $\sim0.1-0.3$ Gyr. They find that post-starburst phases of order $\sim1$ Gyr are only achievable under rare circumstances. The observability of the post-starburst phase is affected by many factors such as merger mass ratio, star formation history, dust content, and strength of feedback processes. For example, the lifetime may be extended if nuclear feedback clears the region of gas and dust, revealing the A stars. Snyder \etal~note a caveat that they only test one feedback model. It is possible that a model of stronger feedback might reduce the sensitivity of the observability on the initial parameters, by having a stronger effect on the burst duration. Feedback models themselves continue to be an open area of investigation \citep[\eg][]{Bourne14} \\

Other simulations of merging galaxies by \citet{Bekki05} use stellar population synthesis code to study post-starburst galaxies. They find that a range of galaxy interactions can produce spectra consistent with post-starbursts and that the initial parameters of the interactions tend to have an effect on the resulting kinematics. Bekki \etal~predict that post-starbursts in elliptical galaxies are likely to have young populations that are more strongly rotating and have lower central velocity dispersions than the respective older population. Investigating these scenarios requires disentangling the young and old populations in observed spectra.

\citet{Stickley14} perform simulations of galaxy mergers that include dissipative gas, and feedback processes from both star formation and nuclear activity. By tracking the stellar velocity dispersion throughout the merger process, they are able to relate \ssig~with other features of the mergers such as star formation and nuclear activity. Stickley \& Canalizo find that the merging galaxies go through several star formation stages, including star formation in the tidal features during the early merger stages to a nuclear starburst that occurs in the later stages during nuclear coalescence. They observed that velocity dispersions as measured from the youngest stars tended to be smaller than that of the total population, and posited this as the result of formation from tidal features. They note that as the merger progresses, those populations become more mixed and the \ssig~values rise to the systemic value. Stickley \& Canalizo also find that a nuclear starburst occurs during the final stages of nuclear coalescence, which is soon followed by the highest SMBH accretion rates of the merger process. While post-starburst galaxies do not have significant nuclear activity, we may be catching these objects before or between episodes of nuclear activity and also sometime late in the coalescence process. Furthermore, it is not impossible to find galaxies that show both nuclear activity and a post-starburst stellar population \citep{Cales13,Cales11}.

\citet{Hiner12} have measured stellar velocity dispersions, \ssig, and black hole masses in a sample of ``post-starburst quasars'' \citep{Cales13,Cales11}. They investigate the \msig, \mlum, and Faber-Jackson relations using these objects and found that some of the systems could be dynamically peculiar. Several of the objects showed some offset from the Faber-Jackson relation. This is consistent with findings by other authors who place quiescent post-starburst galaxies on the Faber-Jackson relation \citep{Swinbank12,Norton01}. However, it is unclear what drives any observed offset. While \citet{Norton01} found no significant correlation between the host luminosity and the young population \ssig, the older population was correlated with host luminosity but offset from the Faber-Jackson relation.

Observations with integral field units (IFU) are opening a new window into the investigation of post-starburst dynamics. \citet{Emsellem07} define a new parameter $\lambda_{\rm R}$ that characterizes the projected stellar angular momentum per unit mass, and they separate ellipticals into ``slow'' and ``fast'' rotators. This technique can be applied to the separate stellar populations of post-starburst galaxies to gain insight into the formation mechanisms of the burst (young) population. \citet{Pracy09} used IFU data to study 10 local post-starburst galaxies, and find that many of them are fast rotators.  \citet{Swinbank12} also use IFU data to investigate 11 post-starburst galaxies. In addition to classifying 10/11 of their objects as fast rotators, they find that the A star population is extended, covering around 33\% of the host galaxy images, indicating that the starburst occurs on galactic scales and is not constrained to the nuclear region. They also find that there are localized star forming regions within the galaxy, not necessarily coinciding with the populations of A stars. A follow-up study by \citet{Pracy13} found more fast rotators and showed that the young stellar population was centrally concentrated in the hosts. \\

With this paper, we describe an investigation into post-starburst galaxies with two main goals. Our first goal is to demonstrate if and how the kinematics of post-starburst populations can be distinguished from those of the underlying older populations. We seek to determine if the two populations exhibit similar kinematics. The results can help test galaxy merger models by determining if post-starburst populations are virialized or if they have peculiar kinematics relative to the older populations. Furthermore, this study is complementary to that of \citet{Hiner12} by establishing context for \ssig~measurements from the Balmer region of the optical spectrum.

We investigate the kinematics of the young and old populations using three separate spectral regions. These regions are commonly used to measure stellar velocity dispersions: the Balmer region (including the Balmer series, Ca H+K lines, and the G band), the \ion{Mg}{1b} region (including H$\beta$), and the Ca Triplet (Ca T) region. The Balmer region, in particular, exhibits many stellar absorption features from both young and old populations in high signal-to-noise spectra. Whereas, the \ion{Mg}{1b} and Ca T regions are dominated by more intermediate or older aged populations.

Our second main goal is to determine what can be learned about the nature of post-starburst galaxies, especially by using reliable measures of the velocity dispersion. To do this, we draw on insights from the Faber-Jackson Relation \citep{FJ76,Desroches07,Nigoche10}, the Fundamental Plane \citep{DD87, HB09}, and we further perform a case-study on a galaxy within our sample. We investigate whether the internal kinematics of the galaxy are consistent with a pressure-supported or rotationally dominated system, and we determine the spatial extent of the younger post-starburst population within the galaxy.

In the following section we describe our sample and data reduction. In Section \ref{fitting} we describe the direct fitting method used on the three spectral regions. In Section \ref{results} we describe the results, comparing the velocity dispersions as measured by different spectral regions and by different population templates. We also investigate the location of the objects in the \ssig-luminosity plane and the Fundamental Plane to see if they conform to the established relations. In Section \ref{case} we present a case study of the internal kinematics of one object in our sample that we observed with the longslit along the major axis of the galaxy. Finally, we summarize and discuss our results in Section \ref{conclude}. Throughout this paper we use the cosmological parameters H$_{0} = 71$, $\Omega_{\rm M} = 0.27$, $\Omega_{\rm vac} = 0.73$, and the cosmological calculator by \citet{Wright06}.

\section{Sample and Observations}

\begin{figure}
\epsscale{1.0}
\plotone{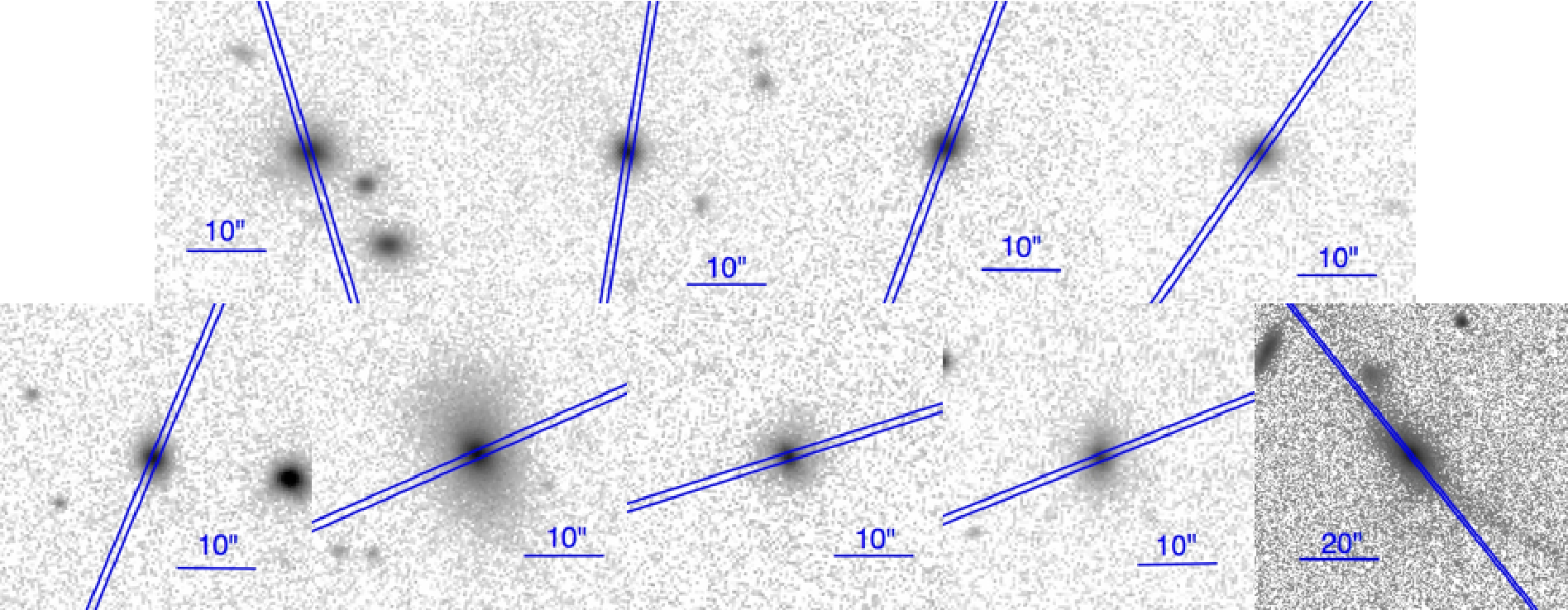}
\caption[Images of the post-starburst galaxy sample]{SDSS $g$ band images of the post-starburst galaxies. The slit used during the LRIS observation is over-plotted in blue and oriented along the observed PA. The objects are ordered by increasing RA {\it top}: No. 1, No. 2, No. 3, and No. 4 {\it bottom}: No. 5, No. 6, No. 7, No. 8, and No. 9. All objects are displayed with a log brightness scale. The brightness of object No. 9 (SDSS $2337-1058$) has been modified to display a faint surface brightness tidal tail.}
\label{img}
\end{figure}

The catalogue of post-starburst galaxies presented by \citet{Goto07} contains 564 galaxies selected from the SDSS DR5 \citep{Adelman-McCarthy07} with the following criteria: H$\delta$ EW $> 5$ \AA, H$\alpha$ EW $> -3.0$ \AA, and [\ion{O}{2}] EW $> -2.5$ \AA. These selection criteria eliminate galaxies with strong emission from either star formation or an active nucleus, but retain galaxies with young, less than $\sim$ 1 Gyr, stellar populations. We selected nine objects from this sample, giving preference to bright objects that best fit into our observing program and could be observed at low airmass. Our selections may be representative of the brighter objects in the parent sample, with no other particular biases. We observed the nine post-starburst galaxies along with several stars (to form host galaxy templates) with the Low Resolution Imaging Spectrograph (LRIS) on Keck 1 during the night of 25 September 2011. Weather conditions were excellent and seeing was $\sim0\farcs6$. Table \ref{observations} shows the objects' redshifts, SDSS $g$ band magnitudes, position angles of the slit during the observation, and the aperture extraction widths. Figure \ref{img} shows the SDSS $g$ band image of each object with the slit orientation over-plotted with width equal to 1\arcsec.

We reduced the data using standard procedures in IRAF. Bias levels and structure were subtracted using the overscan region. Flat field frames were observed using internal halogen lamps. We removed cosmic rays from the 2D spectra using the L.A.\ Cosmic routine \citep{vDokkum01}. We derived a wavelength solution using Hg, Ne, Cd, Ar, and Zn arc lamps. Because of the redshift of the observed galaxies, the dichroic (5600 \AA) divides the \ion{Mg}{1b} region (rest frame $4600 - 5600$ \AA) between the blue and red side LRIS chips. However, the \ion{Mg}{1b} region of the template stars at redshift zero fell solely on the blue side of the instrument. Wavelength calibration errors in the \ion{Mg}{1b} region between the template stars and the galaxies were corrected in the fitting routine using a low-order Legendre polynomial.

Noise in the spectra is dominated by the sky level, which we measured from the wavelength calibrated 2-D frames and for which we assumed Poisson statistics. These measurements were consistent with the RMS measured from the sky-subtracted frames. The spectra we obtained have signal-to-noise ratios ranging from $50-300$, depending on the individual object's brightness and exposure time. The signal-to-noise ratios were calculated using rest-frame wavelength regions $4400-4450$ \AA, $5100-5140$ \AA, and $8100-8175$ \AA~when available.  

After removal of background sky levels, we extracted spectra of the central components of the galaxy using aperture sizes set to the de Vaucouleurs radius as measured by the SDSS DR8 \citep{Aihara11}. In all cases, the SDSS DR8 reports that de Vaucouleurs profiles were better fits than the exponential profiles, indicating that the bulge component of the post-starburst galaxies is dominant over any potential disk component. Furthermore, all objects have sizes larger than the 0\farcs6 seeing limit.

We performed flux calibration using four spectrophotometric standards observed throughout the night. Galactic extinction was minimal. Nevertheless, we corrected for it using the maps of \citet{Schlegel98} and the extinction function of \citet{CCM89}. 

We ensured the blue and red side spectra were on the same resolution by broadening the red side spectra with a Gaussian until it matched the lower resolution of the blue side spectrum. This was only necessary for fitting the \ion{Mg}{1b} region ($4600-5600$ \AA), which is split between the blue and red side spectrographs. The Ca T region is covered by the red side spectrograph for both the template stars and the galaxies. Thus, there was no need to match the resolution of these spectra to that of the blue side.

\begin{table*}\centering
\caption{Observations \label{observations}}
\begin{tabular}{cccccccc}
\hline
N & RA & DEC & z & r [mag] & exptime (s) & PA (deg) & aperture (\arcsec) \\
\hline
1 & $01 19 42.10$ & $+01 07 50.8$ & 0.0899 & 17.30 & 600 & 17 & 1.95 \\
2 & $02 24 27.02$ & $-09 37 39.1$ & 0.0883 & 17.69 & 600 & -9 & 0.83\\ 
3 & $03 03 19.22$ & $-08 36 19.0$ & 0.0757 & 17.63 & 600 & -21 & 1.57 \\
4 & $03 28 02.61$ & $+00 45 02.0$ & 0.2020 & 18.46 & 600 & -35 & 1.98\\
5 & $20 41 32.52$ & $-05 13 34.0$ & 0.0621 & 17.43 & 1200 & -22 & 1.48\\
6 & $21 02 58.78$ & $+10 32 59.8$ & 0.0928 & 16.12 & 1200 & 115 & 4.75\\
7 & $23 07 43.41$ & $+15 25 58.1$ & 0.0697 & 17.44 & 1200 & -72 & 2.13\\
8 & $23 28 06.02$ & $+14 46 25.2$ & 0.0690 & 17.87 & 1200 & -69 & 3.55\\
9 & $23 37 12.71$ & $-10 58 00.3$ & 0.0783 & 16.01 & 1200 & 39 & 6.82\\
\hline
\end{tabular}
\tablecomments{We observed nine post-starburst galaxies from the sample of \citet{Goto07} on the night of 25 September 2011 using LRIS on Keck 1. Objects that were observed for 1200 seconds on the blue side had two exposures on the red side to capture two spectral regions for 600 seconds each.}
\end{table*}

\section{Spectral Fitting}\label{fitting}

The direct fitting method that we adopt has been used extensively in the past \citep[\eg][]{Barth02, Wold07}. We previously employed the method to measure the velocity dispersions of post-starburst galaxies hosting AGN \citep{Hiner12} and in red QSOs \citep{Canalizo12}.

We are primarily interested in the velocity dispersions (second moment) of two individual stellar populations: the underlying old population of the host and the younger post-starburst population. Our overall model is composed of two templates that are each allotted a unique velocity offset and velocity dispersion. The stellar velocity dispersion (\ssig) is modeled through a convolution of the template with a Gaussian. We sum the two broadened templates and multiply the result with a low-order Legendre polynomial as follows: 

\begin{equation}
M(x)= \left[T^{y}(x) \otimes G^{y}(x) + T^{o}(x) \otimes G^{o}(x) \right] \times P(x)
\end{equation}

\noindent
where the superscripts $y$ and $o$ delineate the ``young'' and ``old'' templates, respectively, and $x = log(\lambda)$. The low-order Legendre polynomial accounts for differences between the continuum of the templates and the observed galaxy. Such differences can arise from reddening or from non-stellar emission. Unlike for the ``post-starburst quasars" of \citet{Hiner12}, it is not necessary to model an active nucleus component for these galaxies. The \chisq~function comparing our model with the observed data is minimized through the Amoeba routine from Numerical Recipes (written for IDL by W. Thompson).

The \chisq~function requires a measurement of the errors on each datum. As discussed in the previous section, we calculated errors based on the sky level using Poisson statistics. In order to block a particular region from the fit, we artificially set the error bars on those data to be several orders of magnitude higher than the unblocked region. This essentially eliminates the contribution to the \chisq~function from these data. This was only done to block the \ion{Mg}{1b} absorption line (see below) and additionally for two objects that had emission lines in the \ion{Mg}{1b} region.

\subsection{Stellar Population Templates}

We created galaxy templates by mimicking stellar populations of varying ages. We observed template stars with LRIS during previous observing runs in May and September 2010 \citep[see][for details]{Hiner12}. We created seven stellar population templates by combining the template stars with varying proportions. The templates approximate stellar populations of varying ages by comparison with models from \citet{BC03}. The youngest template was created using a single A0 star. The other templates were created using combinations of all of our observed spectral types: A0, A3, A5, F0, F5, F8, G0, G5, G8, K0, K2, K7, M0 and M8. In order to model ``older'' stellar populations, we decreased the relative contribution from early type stars (A and F). The template that simulates the oldest stellar population consists only of the G5 through M8 spectral types. 

The templates approximate stellar populations of ages 202, 286, 453, 508, 718 Myr, 1.015, and 10.0 Gyr, but should not be considered robust determinations of the actual stellar population age. We simply approximate a luminosity-weighted population in order to determine the stellar kinematics. In our fitting routine, we adopt the 10.0 Gyr template to simulate the old population and test the younger aged templates to identify the best fit young population.

\subsection{Fitting regions: Balmer, \ion{Mg}{1b}, and Ca T}

\begin{figure*}
\includegraphics[width=\textwidth]{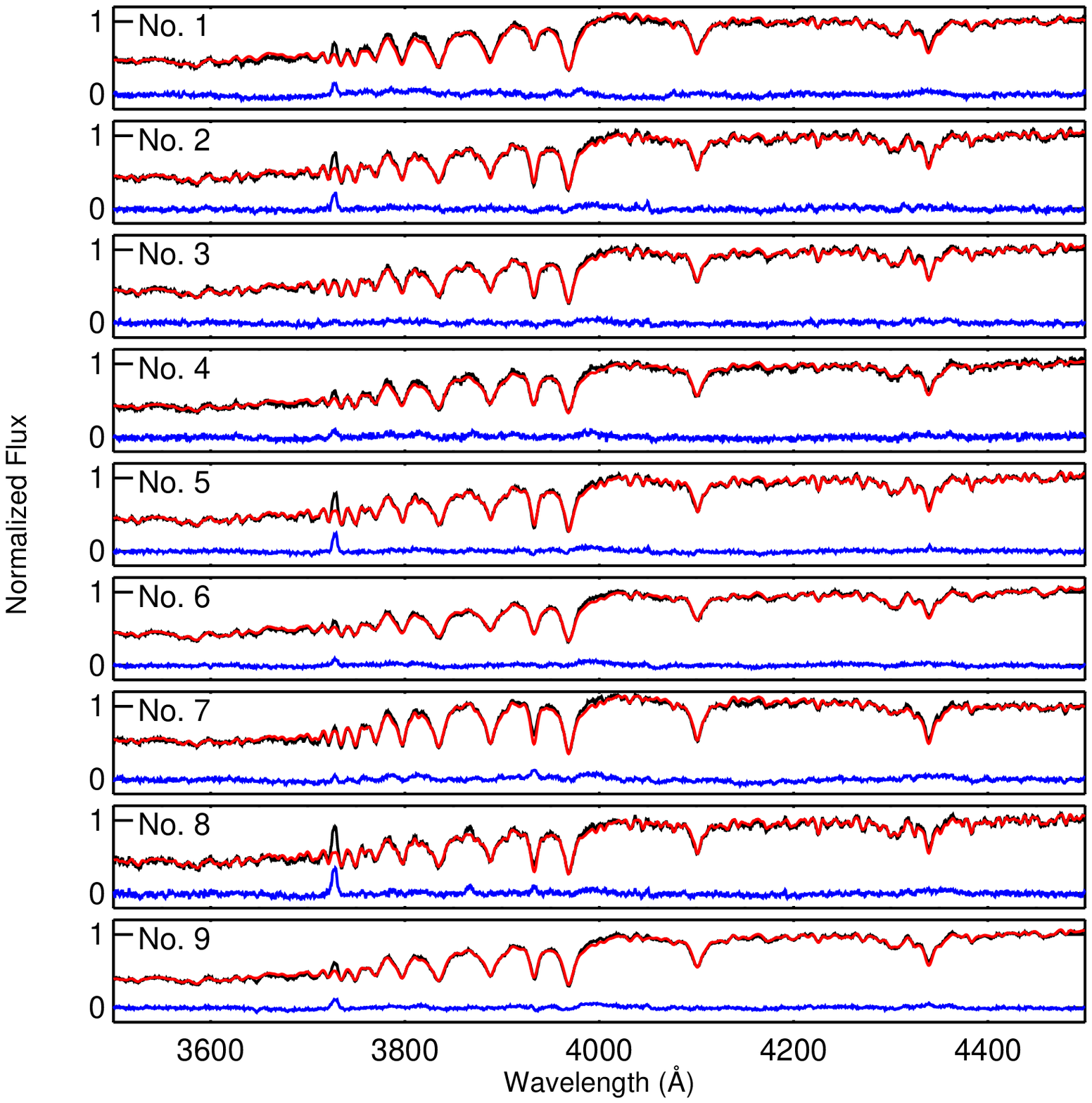}
\caption[Spectral fitting of the Balmer region]{Fits of the Balmer region. Objects are listed in order of increasing RA as presented in Table \ref{observations}. The black line depicts the LRIS spectrum in the rest frame. The best fit model is over-plotted in red. The residuals are plotted in blue. }
\label{balfits}
\end{figure*}

We fit the $3500-4500$ \AA~region of the galaxies, which includes the Balmer series beginning with H$\gamma$, the Ca H+K lines, the G-band, and a host of other weaker spectral features. While this region is dominated by the Balmer series, the weaker spectral features are clear in the high signal-to-noise data. These weaker features can be easily attributed to the later-type stars, and are dominant in the older spectral template. Thus, when we fit this region, the young stars are best constrained by the Balmer series, and the old stars are best constrained by the weaker features, the G-band, and the Ca lines. Figure \ref{balfits} shows the results of the fits to the Balmer region. The data are shown in black, and the best-fit model in red. We show the residuals in blue beneath each fit. 

In the case of object SDSS $2307+1525$ (No. 7), we adopt a model that did not produce the lowest \chisq. The model that produced the lowest \chisq~value did not match the Ca K feature and also 
was dominated by the young population template (286 Myr) with zero contribution from the older template. This being unrealistic, we refit the object using the youngest-aged template (202 Myr) instead of the 286 Myr template. This better matched both the Ca K feature and maintained a contribution from the older population.

We fit the $4600 - 5600$ \AA~region of the spectra, which includes the H$\beta$ absorption line, the \ion{Mg}{1b} line, and other stellar features. The \ion{Mg}{1b} region is typically dominated by the older population. In order to place a strong constraint on the young population contribution, we included the nearest Balmer line in this region. While many have focused on the \ion{Mg}{1b} region alone ($\lambda$ $> 5000$\AA), including the H$\beta$ line is not unprecedented. \citet{BenderNieto90} use a similar region in the study of dwarf spheroidals and compact elliptical galaxies. 

We also note that while the post-starburst galaxy selection criteria generally select against emission line galaxies, some weak emission is permitted in the criteria by \citet{Goto07}. Five of the nine galaxies we study have detectable H$\alpha$ emission, raising the possibility of Balmer line filling. In the worst case scenario, object SDSS $2041-0513$ (No. 5), an H$\beta$ emission line may be filling up to 20\% of the underlying absorption line. Any reddening of the nebular emission would decrease that filling factor. Other galaxies have lower H$\alpha$ emission-to-absorption ratios and have lower contributions to H$\beta$ ($\sim5\%$).  

\begin{figure*}
\includegraphics[width=\textwidth]{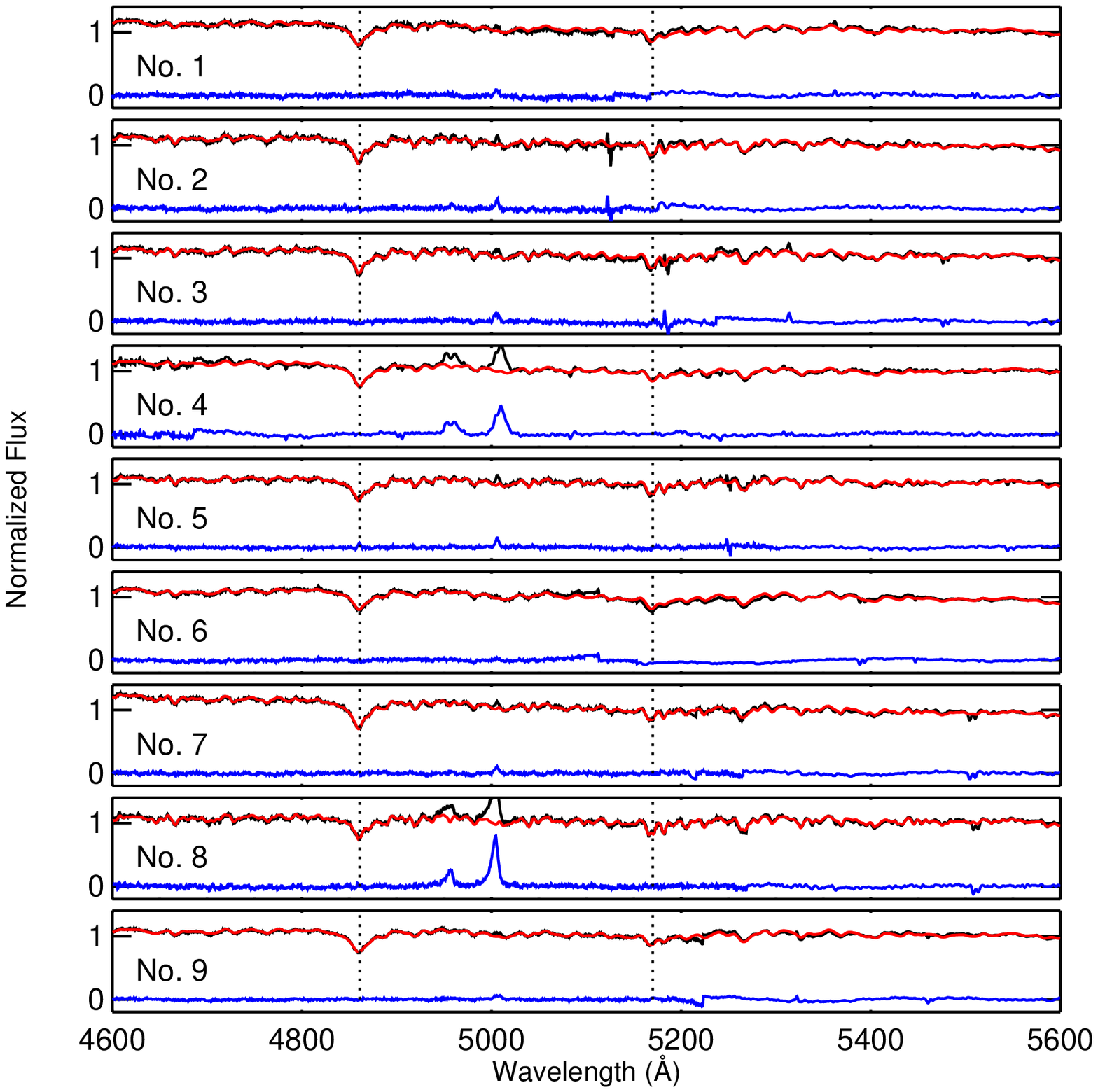}
\caption[Spectral fitting of the \ion{Mg}{1b} region]{Fits of the \ion{Mg}{1b} region. The dotted lines indicate the location of H$\beta$ and \ion{Mg}{1b}. Object order and plotted colors are the same as Fig.~\ref{balfits}}
\label{mgfits}
\end{figure*}

The Ca T region is dominated by light from late type stars, and has been used in the past as a way of measuring velocity dispersions in a variety of galaxy morphological types \citep[\eg][]{Dressler84}. We observed stars with four different spectral types to create templates of the Ca T region: a G0, G5, K0, and K5 star, from which we created three templates. The first consisted of all four stars with weights favoring the G stars. The second had no contribution from the G0 star, and the third consisted solely of the K stars. The contribution from A and F stars in this region is mostly continuum. Because our Legendre polynomial can already mimic the continuum, we did not include early-type stars while fitting the Ca T. Including a separate young population would actually introduce degeneracies in the fit that may not have a physical interpretation. 

\begin{figure}
\epsscale{1.0}
\plotone{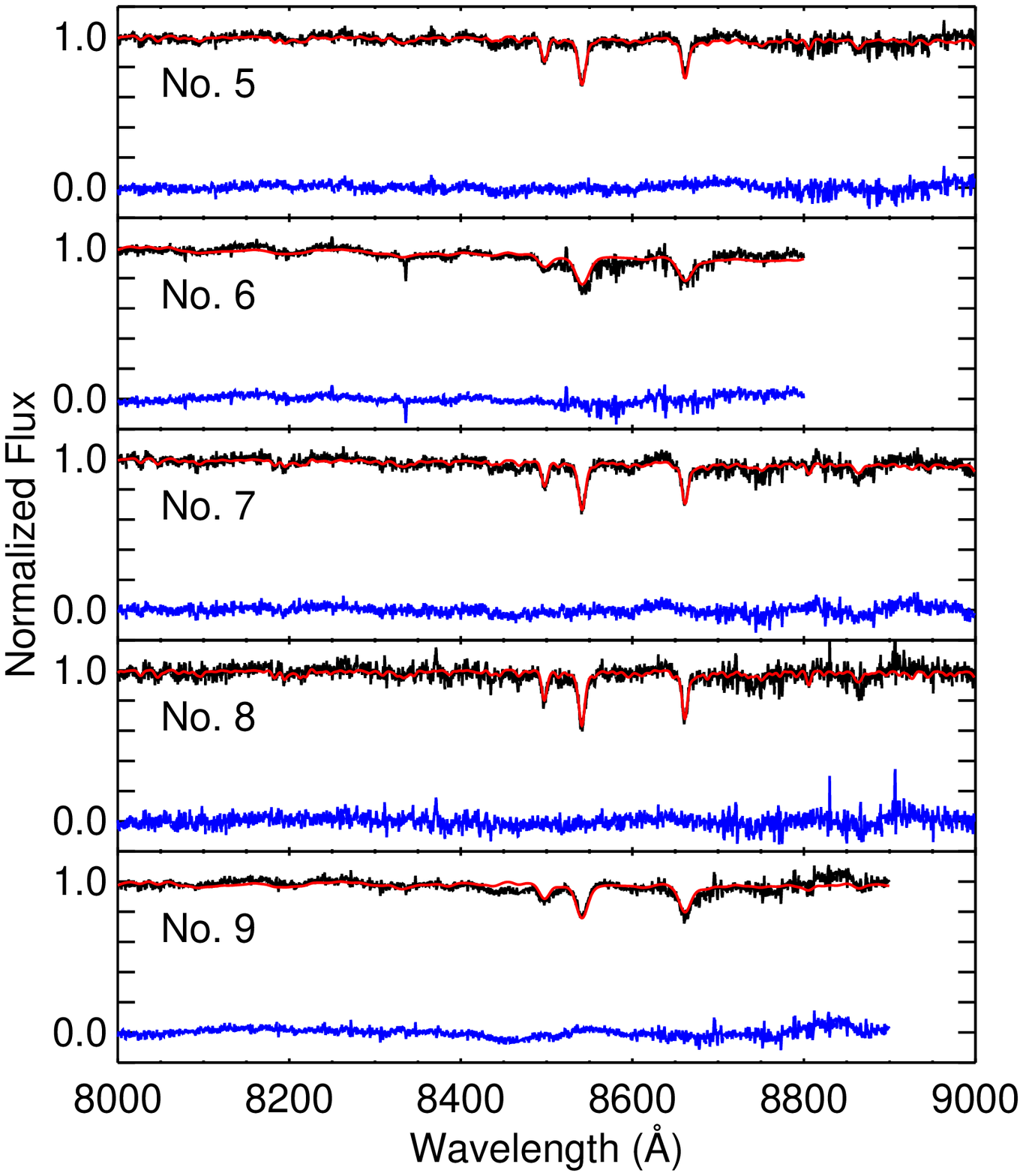}
\caption[Spectral fitting of the Ca T region]{Fits of the Ca T region. The plotted colors are the same as Fig.~\ref{balfits}, and the objects are ordered by increasing RA.}\label{catfits}
\end{figure}

We fit each Ca T region with the three templates and each of the individual stars for a total of seven different galaxy templates. Many of the fits were comparable when inspected by-eye with ranges in \ssig~that match our adopted uncertainties (see below). We chose the ``best-fit'' as the one which produced the minimum \chisq. Figure \ref{catfits} shows the fitting results of the Ca T region.

\begin{figure}
\plotone{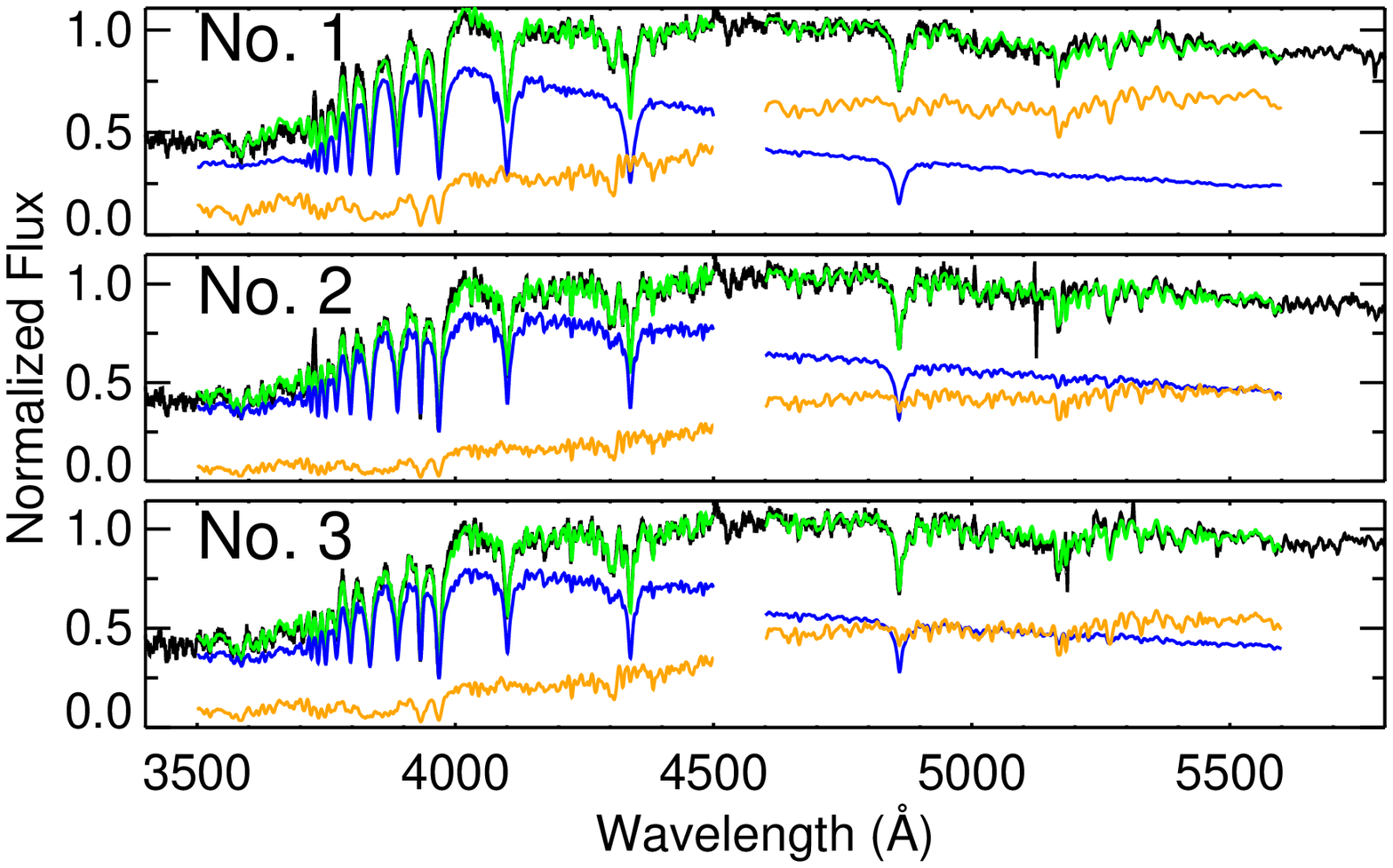}
\caption[Fitted components a]{Fits of the Balmer and \ion{Mg}{1b} regions with components shown. The data are plotted in black and the best fit models over-plotted in green. The relative young and old components are shown in blue and orange, respectively. Objects shown are SDSS $0119+0107$, SDSS $0224-0937$, and SDSS $0303-0836$. }
\label{figure:balmga}
\end{figure}

\begin{figure}
\plotone{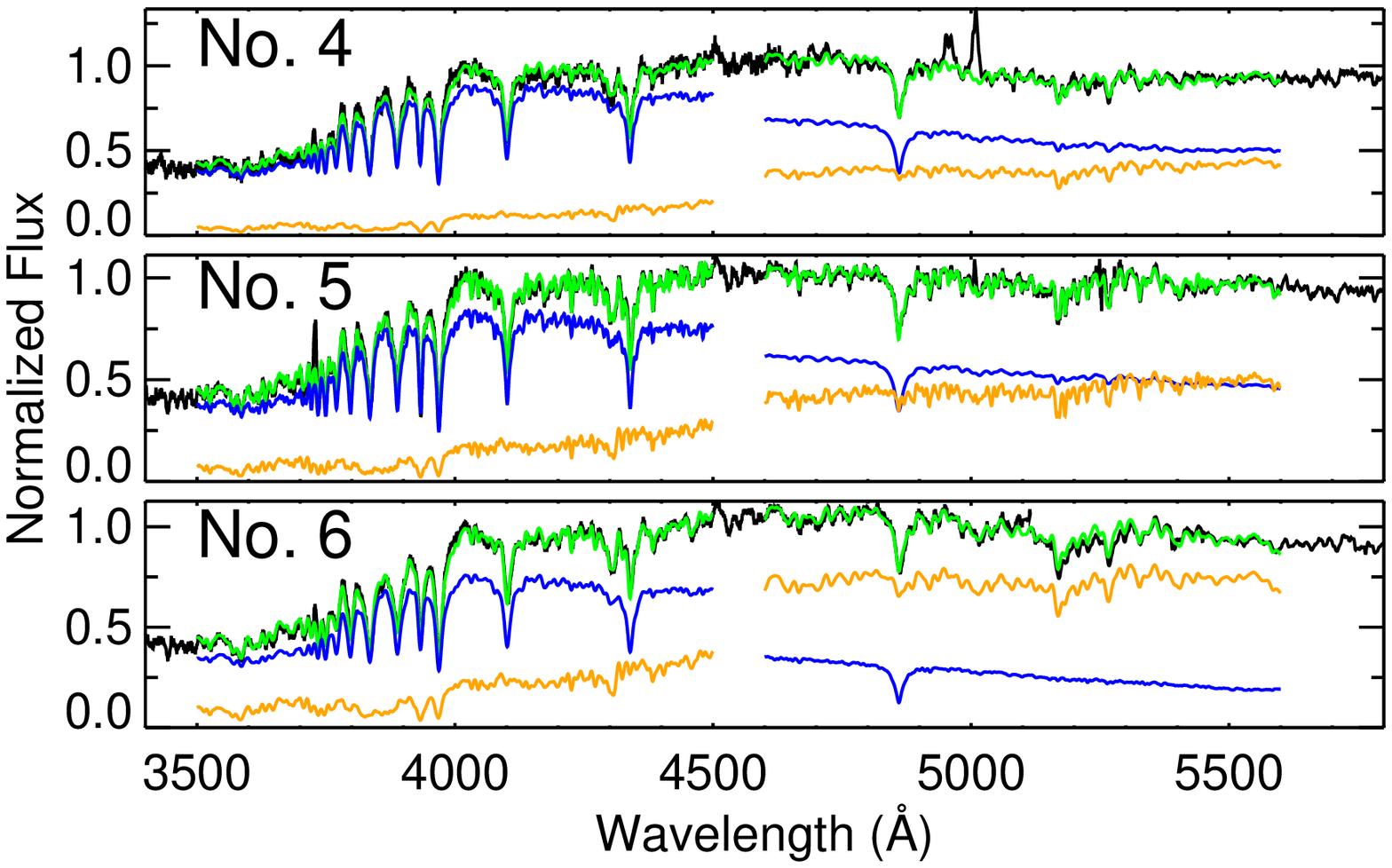}
\caption[Fitted components b]{Fits of the Balmer and \ion{Mg}{1b} regions with components shown. The data are plotted in black and the best fit models over-plotted in green. The relative young and old components are shown in blue and orange, respectively. Objects shown are SDSS $0328+0045$, SDSS $2041-0513$, and SDSS $2102+1032$.}
\label{figure:balmgb}
\end{figure}

\begin{figure}
\plotone{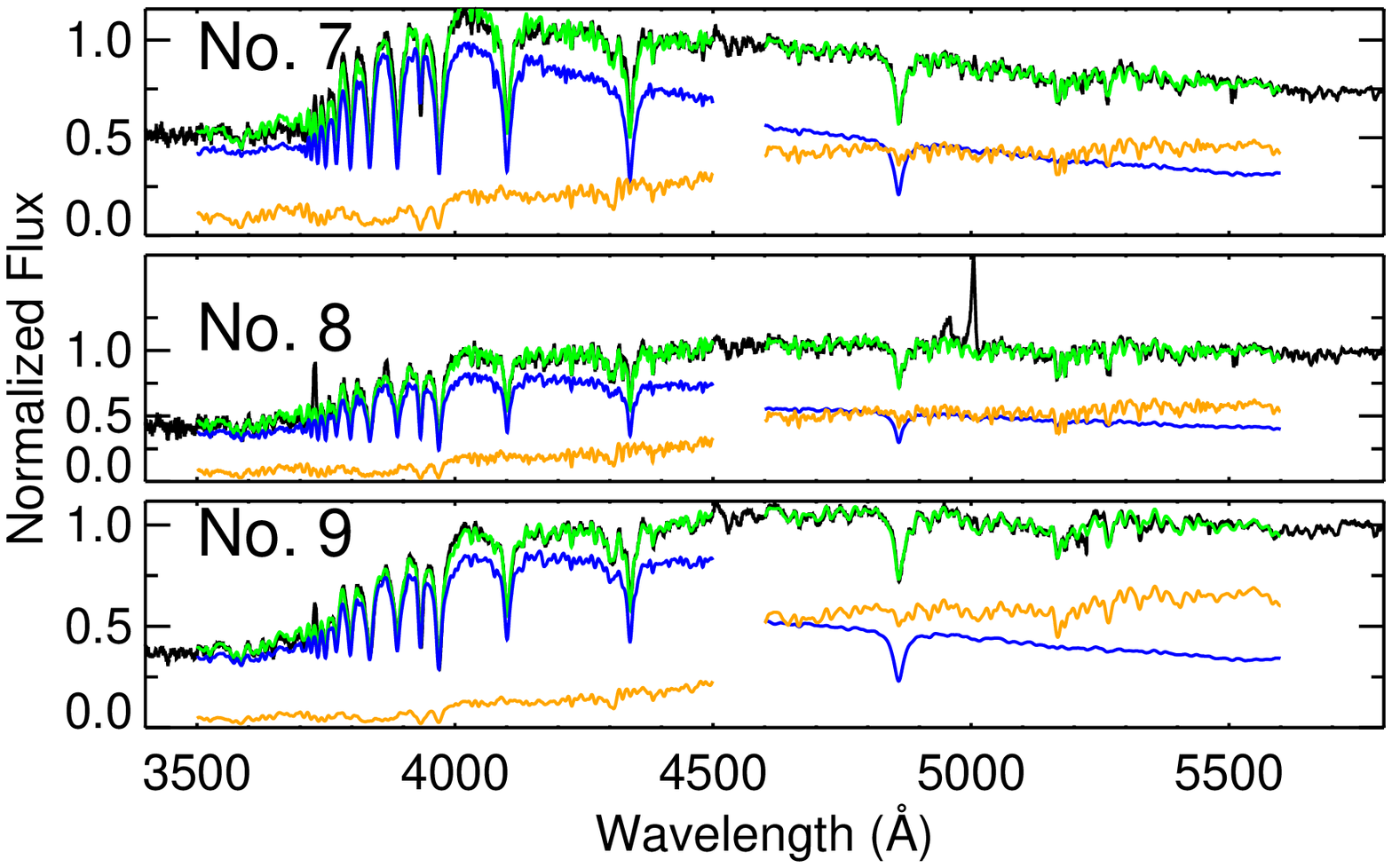}
\caption[Fitted components c]{Fits of the Balmer and \ion{Mg}{1b} regions with components shown. The data are plotted in black and the best fit models over-plotted in green. The relative young and old components are shown in blue and orange, respectively. Objects shown are SDSS $2307+1525$, SDSS $2328+1446$, and SDSS $2337-1058$. }
\label{figure:balmgc}
\end{figure}

In Fig.~\ref{figure:balmga}, \ref{figure:balmgb}, and \ref{figure:balmgc}, we show the best fit models of the Balmer region and the \ion{Mg}{1b} region simultaneously. There is good agreement between the two regions, with the exception of two objects. Objects SDSS $2102+1032$ (No. 6) and SDSS $2337-1058$ (No. 9) were each fit best by the 286 Myr young template in the Balmer region, but in the \ion{Mg}{1b} region they were each best fit with the 202 Myr template. All other objects were fit by the same aged templates between the two regions. We note the presence of some flux variation between the components of different wavelength regions. In the two cases mentioned above (also the worst offenders), this is likely due to mis-matched young stellar components. In the other cases it may be a result of a slight degeneracy between the young component template, which is largely featureless in the \ion{Mg}{1b} region with the exception of the H$\beta$ line, and the multiplicative Legendre polynomial.

\subsection{Uncertainties}

We simulated the effects of noise in the spectra to determine its associated uncertainty. Recall that the signal-to-noise ratio is high, so we do not expect this to be a large factor. We added randomized Poisson noise determined from the sky level to the best-fit models as a basis to create a set of synthetic spectra. Fits of the spectra then reproduce the best-fit model with variations in the parameters that are due to sky noise. We recovered the ``best-fit'' velocity dispersions of the galaxy with errors of only a few km s$^{-1}$, indicating that spectral noise is only a small contributor to the overall uncertainties.

In order to better understand the uncertainties in each wavelength region contributed by template mis-match, we examine the difference in velocity dispersion between the fitted templates. We calculate the standard deviation of velocity dispersions fitted using other templates. Even this method produced some values that were less than 10 km s$^{-1}$. In these cases, because the template mis-match error was of a similar magnitude as the sky noise error, we added the two in quadrature to determine the overall uncertainty in our fits. 

Uncertainty on the old population \ssig~values due to template mismatch tends to be larger than on the young population, especially when measured in the Balmer region. This is most likely because our templates include more contribution from later-type stars when we artificially age them. Thus, when testing the older population templates, a degeneracy arises between the young and old population. The young population is still additionally constrained by the Balmer series, but the older population \ssig~parameter is free to adjust to any inaccuracies generated by including later type stars which are assigned the younger population \ssig.

We also fit each spectrum by adding a single K to a single A star. The K star template is an average of our observed stars that have K spectral types (K0, K2, K5, and K7). Similarly the A star template is an average of our observed A3 and A5 stars. We note that the Ca K line in the A0 spectrum was significantly weaker compared to the A3 and A5 stars, so we tested the A0 star separately. We found that in all objects with the exception of one (SDSS $0119+0107$, No. 1), the average of the A3 and A5 stars produced a result with lower \chisq~than the single A0 star. The difference in reduced \chisq~values was frequently large ($>10$), and for these fits the difference in \ssig~could be by a factor of $\sim2$. We present the results of these fits in Table \ref{ka}, and we discuss the differences between the single-star measurements and the population template measurements in the following section.

For the Ca T region, because we fit the galaxy with only one stellar population template and are interested in only the \ssig~parameter, we were able to estimate the errors using the 95\% confidence interval determined by the \chisq~statistics. This is done by fitting the spectra multiple times with the \ssig~parameter held constant and allowing the other parameters to be freely determined. Each fit produces a \chisq~value with the smallest value occurring at the ``best-fit'' model. The confidence interval is then determined by examining the \ssig~values to the right and left of the best-fit value when the \chisq~is increased by four. 

\begin{table*}\centering
\caption{Velocity Dispersions \label{vdisp}}
\begin{tabular}{cccccccc}
\hline
N & RA & DEC & \multicolumn{2}{c}{Balmer} & \multicolumn{2}{c}{\ion{Mg}{1b}} & Ca T \\
 & & & Young & Old & Young & Old & Old \\
\hline
1 & $01 19 42.10$ & $+01 07 50.8$ & $132\pm10$ & $163\pm57$ & $115\pm97$ & $206\pm29$ & \nodata \\
2 & $02 24 27.02$ & $-09 37 39.1$ & $81\pm9$ & $76\pm44$ & $88\pm10$ & $109\pm20$ & \nodata \\ 
3 & $03 03 19.22$ & $-08 36 19.0$ & $101\pm5$ & $112\pm25$ & $93\pm13$ & $114\pm3$ & \nodata \\
4 & $03 28 02.61$ & $+00 45 02.0$ & $157\pm13$ & $183\pm32$ & $179\pm5$ & $157\pm21$ & \nodata \\
5 & $20 41 32.52$ & $-05 13 34.0$ & $74\pm12$ & $83\pm16$ & $211\pm100$ & $72\pm12$ & 109$^{+24}_{-22}$ \\
6 & $21 02 58.78$ & $+10 32 59.8$ & $195\pm18$ & $174\pm32$ & $89\pm13$ & $227\pm5$ & 248$^{+42}_{-55}$ \\
7 & $23 07 43.41$ & $+15 25 58.1$ &  $86\pm9$ & $111\pm51$ & $137\pm13$ & $103\pm30$ & 101$^{+23}_{-21}$ \\
8 & $23 28 06.02$ & $+14 46 25.2$ & $75\pm11$ & $33\pm36$ & $150\pm9$ & $61\pm15$ & 81$^{+19}_{-17}$ \\
9 & $23 37 12.71$ & $-10 58 00.3$ & $150\pm8$ & $167\pm17$ & $224\pm28$ & $189\pm9$ & 206$^{+44}_{-42}$ \\
\hline
\end{tabular}
\tablecomments{Best fit velocity dispersions (km s$^{-1}$) for the young and old components of each galaxy. The young population templates used in the Balmer region for all objects correspond to an approximate age of 286 Myr with the exceptions of No. 1 (SDSS $0119+0107$) and No. 7 (SDSS $2307+1525$), for which we used a template corresponding to 202 Myr.}
\end{table*}

\begin{table*}\centering
\caption{K+A tests \label{ka}}
\begin{tabular}{ccccccc}
\hline
N & RA & DEC & Balmer & Balmer & \ion{Mg}{1b} & \ion{Mg}{1b} \\
 & & & A & K & A & K \\
\hline
1 & $01 19 42.10$ & $+01 07 50.8$ & $135\pm10$ & $176\pm57$ & $132\pm97$ & $223\pm29$ \\ 
2 & $02 24 27.02$ & $-09 37 39.1$ & $105\pm9$ & $103\pm44$ & $123\pm10$ & $101\pm20$ \\ 
3 & $03 03 19.22$ & $-08 36 19.0$ & $132\pm5$ & $121\pm25$ & $140\pm13$ & $103\pm3$ \\ 
4 & $03 28 02.61$ & $+00 45 02.0$ & $183\pm13$ & $195\pm32$ & $236\pm5$ & $163\pm21$ \\ 
5 & $20 41 32.52$ & $-05 13 34.0$ & $102\pm12$ & $97\pm16$ & $248\pm100$ & $81\pm12$ \\ 
6 & $21 02 58.78$ & $+10 32 59.8$ & $220\pm18$ & $191\pm32$ & $113\pm13$ & $240\pm5$ \\
7 & $23 07 43.41$ & $+15 25 58.1$ & $133\pm9$ & $88\pm51$ & $155\pm13$ & $75\pm30$ \\ 
8 & $23 28 06.02$ & $+14 46 25.2$ & $113\pm11$ & $46\pm36$ & $166\pm9$ & $66\pm15$ \\ 
9 & $23 37 12.71$ & $-10 58 00.3$ & $179\pm8$ & $170\pm17$ & $213\pm28$ & $183\pm9$ \\ 
\hline
\end{tabular}
\tablecomments{Best fit velocity dispersions (km s$^{-1}$) for the using K+A templates. The K star is an average of K0, K2, K5, and K7 stars. The A star is an average of an A3 and A5 star. Using a single A0 star fit object No. 1 (SDSS $0119+0107$) better than the A star average in both the Balmer and \ion{Mg}{1b} regions. We adopt uncertainties for these fits that are identical to those presented in Table \ref{vdisp}, which are determined by quantifying the template mis-match.}
\end{table*}

\section{Results}\label{results}

In this section we present the main results of our study. Recall our first objective is to determine the feasibility and reliability of measuring kinematics of individual young and old stellar populations in post-starburst galaxies. We present the \ssig~values from the two populations in the different wavelength regions. We also measure the \ssig~values using single spectral types (``K+A") and compare those with the values found using the population model templates. We examine the object luminosities and sizes and plot them on the Faber-Jackson Relation and Fundamental Plane in order to determine if the post-starburst galaxies are kinematically peculiar compared to others.

\subsection{Velocity Dispersions}

As discussed above, we have fit both young and old populations with individual \ssig~in both the Balmer and the \ion{Mg}{1b} regions. We additionally fit the late-type (likely old population) stars in the Ca T region when available. Here we show the differences in velocity dispersions as measured from the three different wavelength regions and two populations. In order to determine if measurements are consistent with each other, for example between measurements of the same population from different wavelength regions, we calculate a parameter, $\delta$, as follows:

\begin{equation}
\delta = \frac{Y - X}{\sqrt{\sigma_Y^2 + \sigma_X^2}} 
\end{equation}

\noindent
where Y represents the ordinate and X the abscissa values, and the $\sigma$ values are the respective errors. If this value falls within the range $-1$ to $1$, the Y measurement is considered consistent with the X measurement within the error.

We also perform linear least-squares fits of the data for each comparison using the fitexy task in IDL, which accounts for errors in both the ordinate and abscissa values. If the measurements were perfectly consistent, we expect to find a one-to-one correlation between them, i.e., a slope of 1 in the least-squares fit. Table \ref{fitparams} contains a summary of the fitted parameters, including the intercept, slope, uncertainties on fitted parameters, \chisq~values, and the \chisq~probability value, q. The q-value is to be interpreted as the likelihood that a correct model would produce a \chisq~value at least as large as the value determined in our fit. In the following paragraphs, we describe the results of comparing the \ssig~values measured by different methods.

\begin{figure*}
\epsscale{1.0}
\begin{tabular}{cc}
\epsfig{file=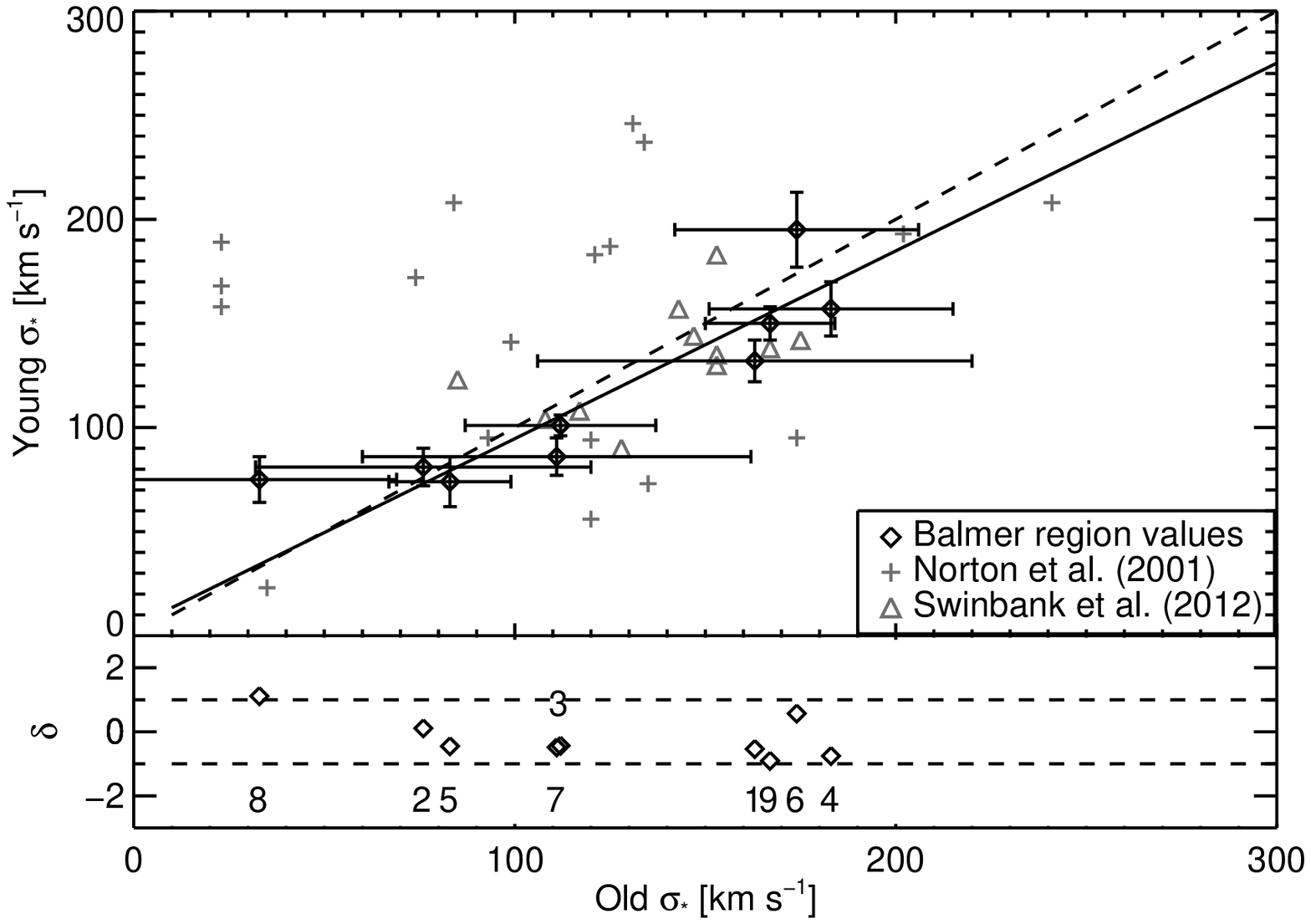, width=0.5\linewidth,clip= } & \epsfig{file=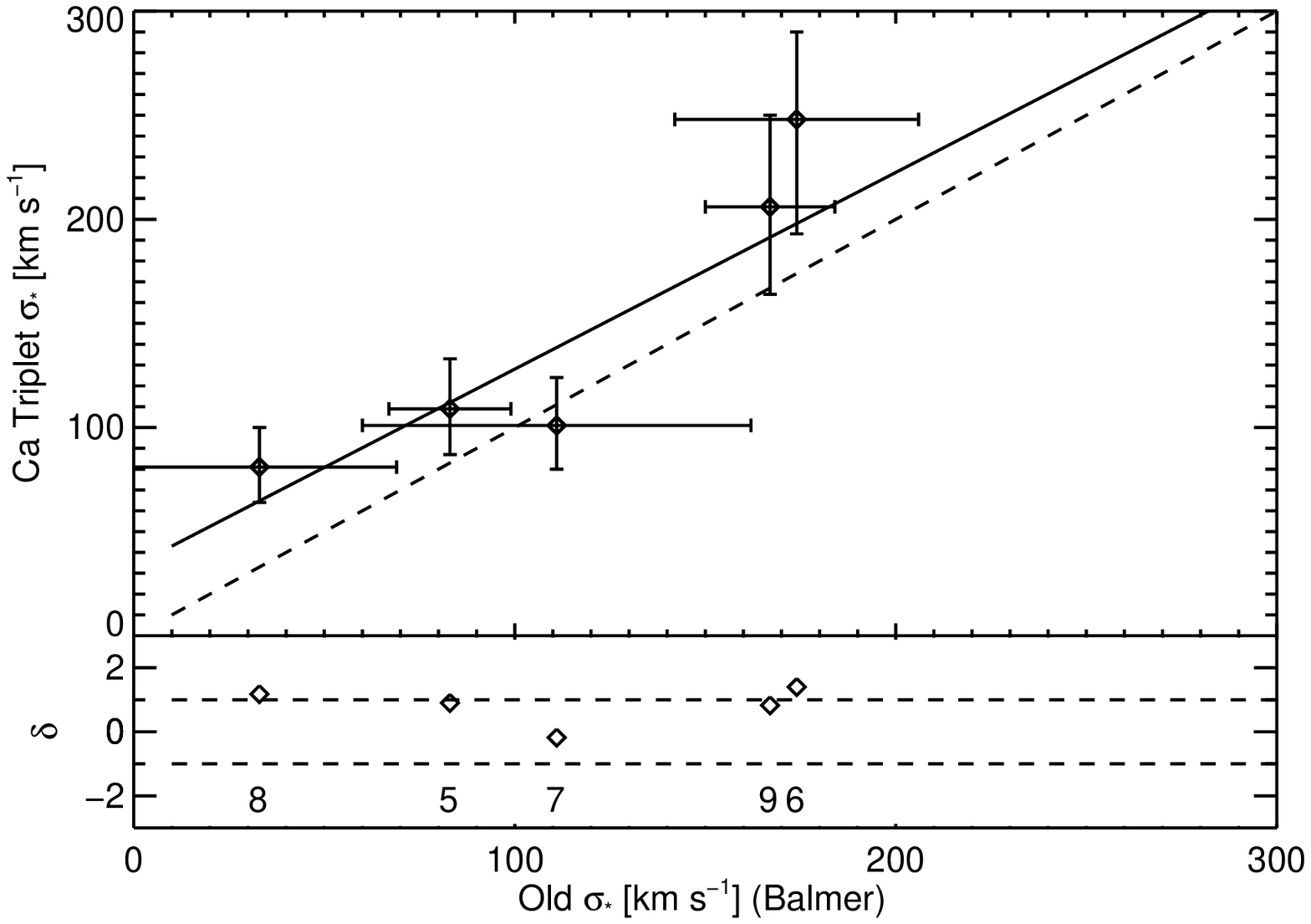, width=0.5\linewidth,clip= }\\
\epsfig{file=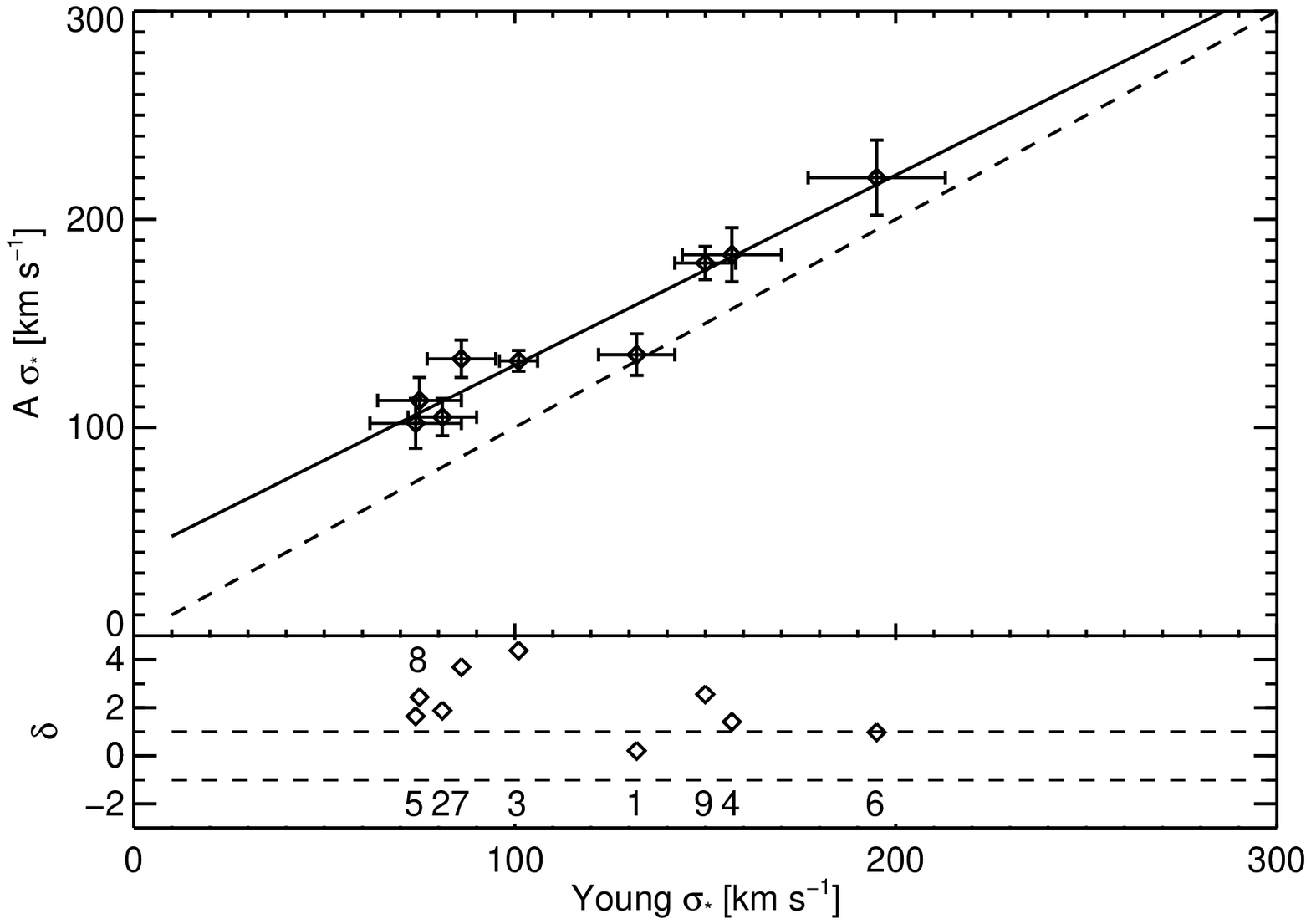, width=0.5\linewidth,clip= } & \epsfig{file=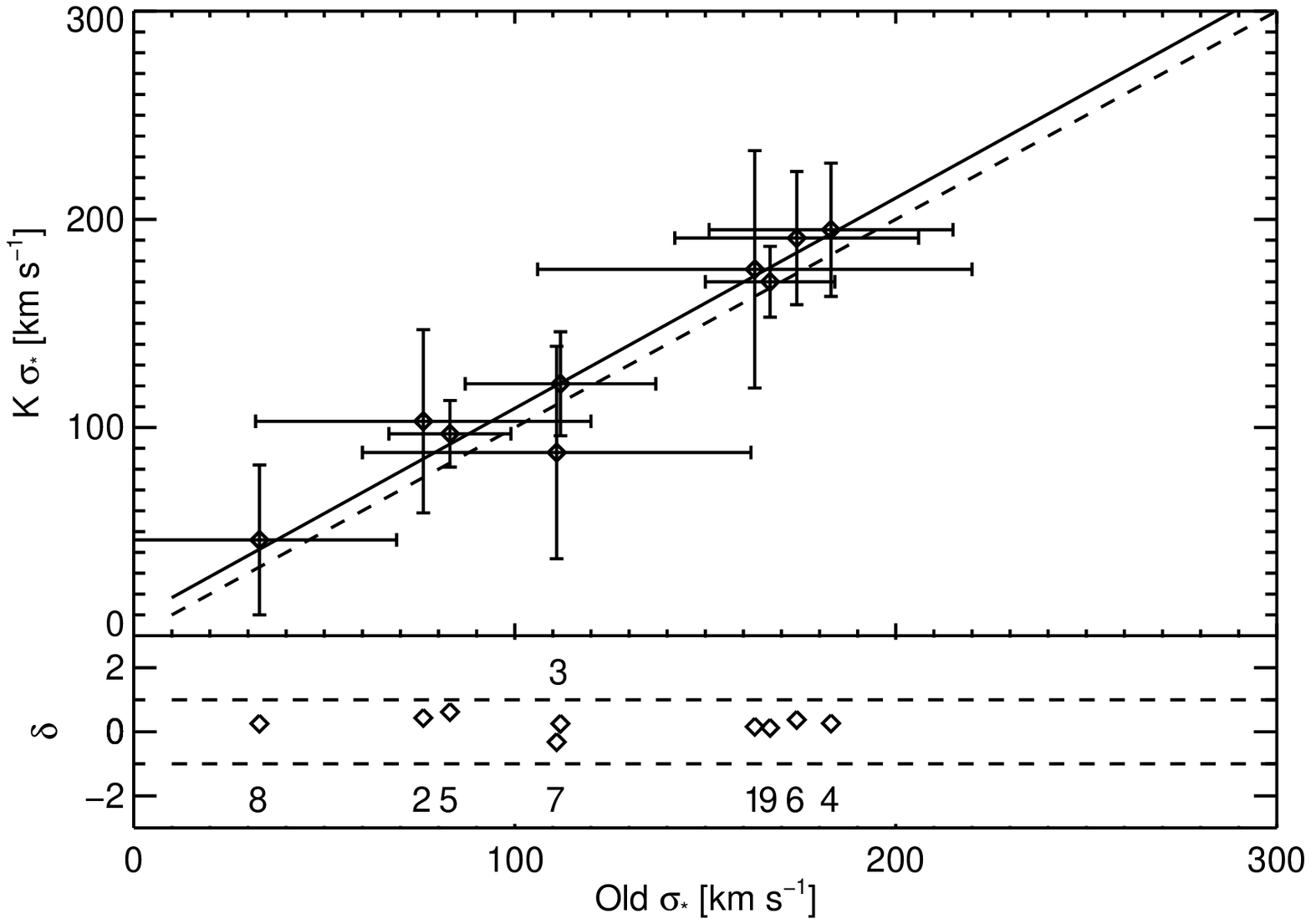, width=0.5\linewidth,clip= }\\
\end{tabular}
\caption{Velocity dispersion comparisons for measurements taken in the Balmer region. In each panel the dotted line has slope of one, and the solid line is a fit to the data. The fitted parameters are shown in Table \ref{fitparams}. In the top left panel, the \ssig~of the young population is compared to that of the old population. Comparison objects from \citet{Norton01} and \citet{Swinbank12} are shown, but not included in the fit of the correlation. The \ssig~measurements from the Ca T region are compared to the old population (Balmer region) in the top right panel. The panels on the bottom compare \ssig~as measured with average A and K star templates with the those of the respective populations they are intended to mimic. All of the fitted correlations are consistent with a one-to-one relationship.}
\label{baldisp}
\end{figure*}

The top left panel of Fig.~\ref{baldisp} shows the \ssig~of the young components compared to the old components as measured in the Balmer region. The dashed line is a one-to-one correlation, and is the same for the other figures that follow in this subsection. The objects are correlated and roughly match the one-to-one line with some scatter. The solid line is a linear fit to the data, which shows a slope that is close to unity ($0.90\pm{0.22}$). This slope is dependent on the number of included points and would be more robustly determined with a more statistically significant sample. With the exception of one object, all of the young population \ssig~are consistent with the old population \ssig~within the estimated error bars. On the surface, this may indicate the that the kinematics of the young and old populations are determined by the same potential. We also show comparison objects from \citet{Norton01} and \citet{Swinbank12} in this panel. As discussed below, these authors use different methods for measuring \ssig~than we do, and thus these objects are not included in the determination of the correlation between young and old populations.

The top right panel of Fig.~\ref{baldisp} compares the old velocity dispersion from the Balmer region with the velocity dispersion measured from the Ca T region. If the G-band and weaker stellar absorption lines in the Balmer region are produced by the same stellar population that produces the Ca T absorptions, then we might expect the velocity dispersions to be the same within the measurement error. Indeed the two are correlated (with a linear slope of $0.95\pm{0.25}$), but it appears that the Ca T dispersions are systematically offset from the Balmer region dispersions by approximately $34\pm{35}$ km s$^{-1}$. While there is significant uncertainty in this result due to the estimated errors, it is interesting, and we will return to in the discussion section.

\begin{figure*}
\epsscale{1.0}
\begin{tabular}{cc}
\epsfig{file=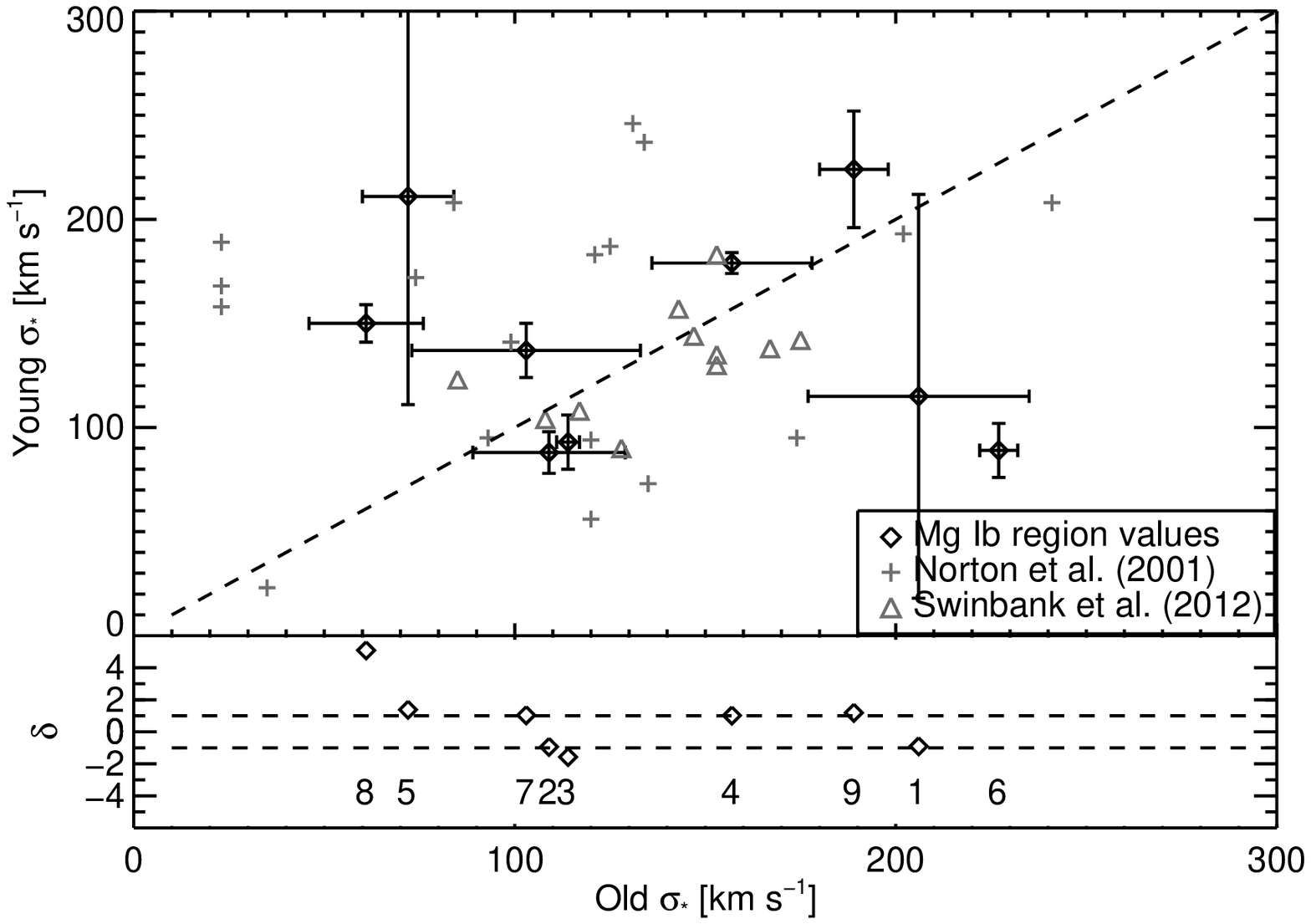, width=0.5\linewidth,clip= } & \epsfig{file=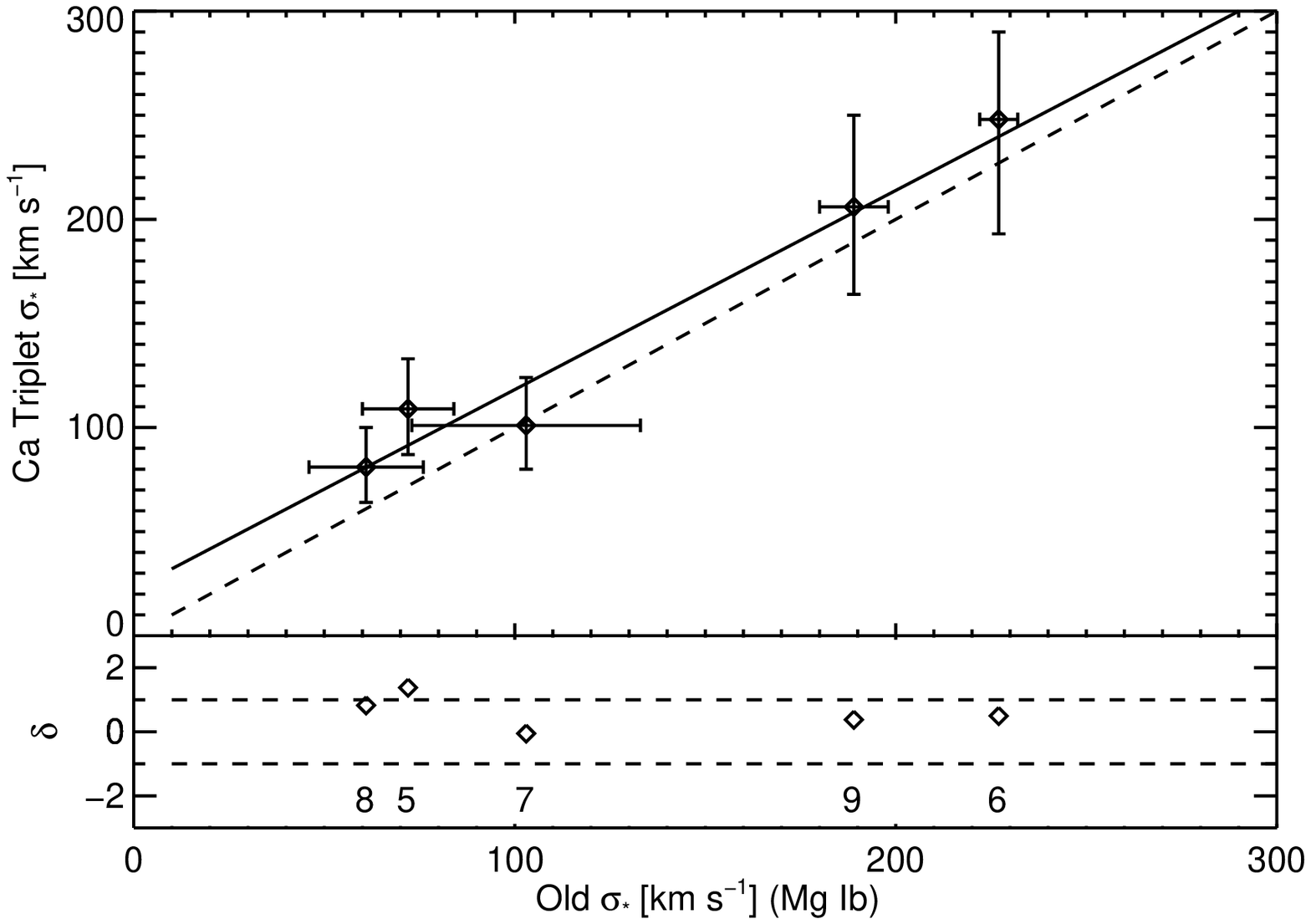, width=0.5\linewidth,clip= }\\
\epsfig{file=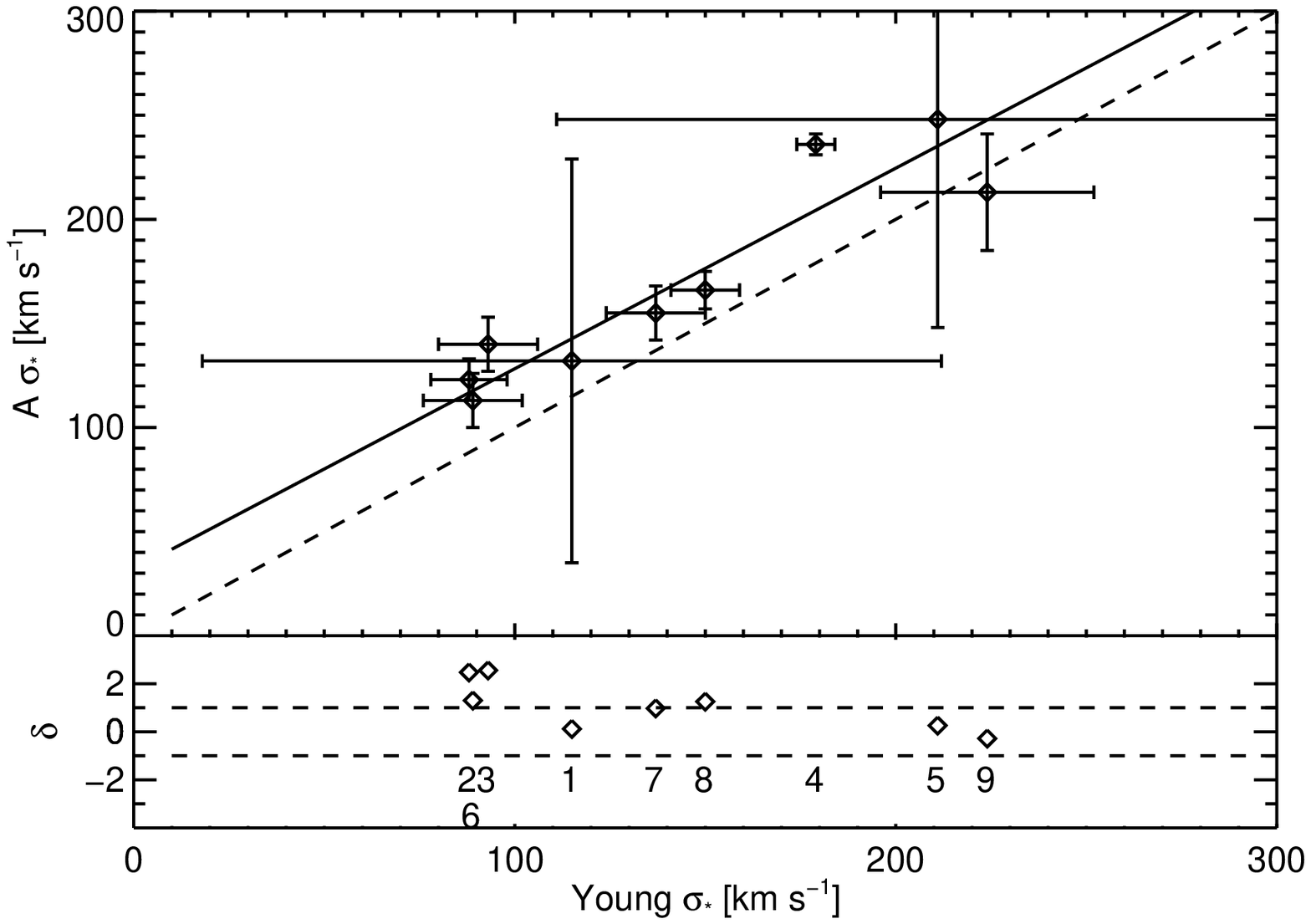, width=0.5\linewidth,clip= } & \epsfig{file=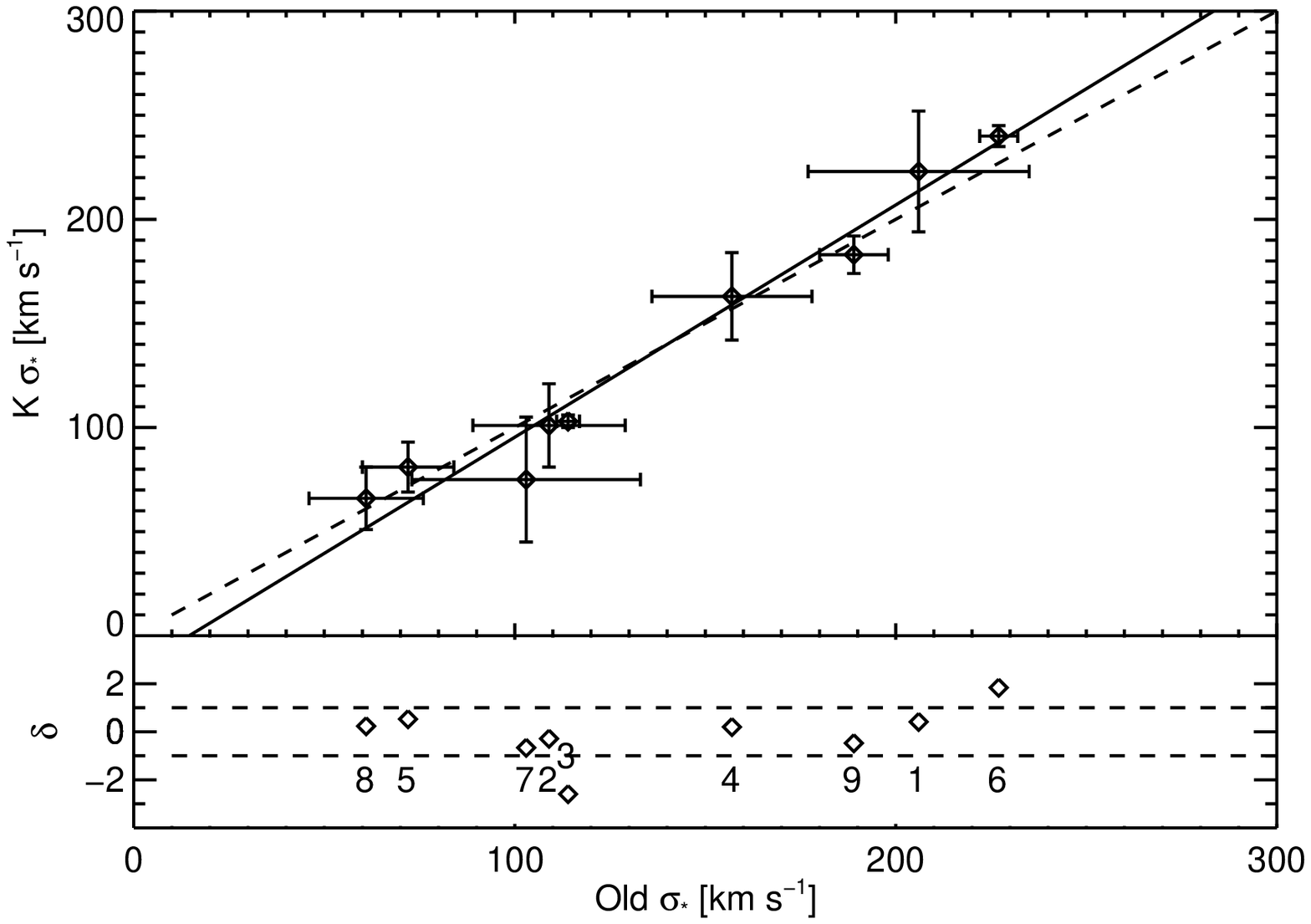, width=0.5\linewidth,clip= }\\
\end{tabular}
\caption{Velocity dispersion comparisons for measurements taken in the \ion{Mg}{1b} region. In each panel the dotted line has slope of one, and the solid line is a fit to the data. The fitted parameters are shown in Table \ref{fitparams}. In the top left panel, the \ssig~of the young population is compared to that of the old population. Comparison objects from \citet{Norton01} and \citet{Swinbank12} are also shown (same as Fig.~\ref{baldisp}). The \ssig~measurements from the Ca T region are compared to the old population as measured in the \ion{Mg}{1b} region in the top right panel. The panels on the bottom compare \ssig~as measured with average A and K star templates with the those of the respective populations they are intended to mimic. No correlation is seen between the young and old population \ssig~measurements in the top left panel, while all other correlations are consistent with a one-to-one relationship.}
\label{mgdisp}
\end{figure*}

The bottom panels of Fig.~\ref{baldisp} compare the fits of the Balmer region using our stellar population models with the simpler method of using a K+A star model. We remind the reader that we tested an average template composed of the A3 and A5 stars, as well as the single A0 star. For this comparison we adopt symmetric error bars. That is, the errors for the K+A measurements that are equivalent to those produced by using the respective young or old population templates. The reason for this is that the errors we determined for the population models are intended to quantify the template mis-match uncertainty, and here we are essentially testing the intentional mis-match between the K and A star templates from the old and young populations, respectively. The error on the \ssig~measurements from the K and A star templates is the template mis-match error.

The bottom left panel of Fig.~\ref{baldisp} shows the young component compared to the single A star model.  In all cases but one the A star over estimates \ssig~by approximately $38\pm{16}$ km s$^{-1}$. The fitted correlation has slope of $0.91\pm{0.13}$. The single point where the A star \ssig~is the same as that of the stellar population represents object SDSS $0119+0107$ (No. 1). The population template that best fit this object was simulated with a single A0 star, so in fact the two templates use the same spectrum to model the young component. The bottom right panel of the Fig.~\ref{baldisp} compares the old component population model measured in the Balmer region with the single K star model. Here we find that the correlation is close to one-to-one with a slope of $1.0\pm{0.10}$ and an offset of only $8\pm{13}$ km s$^{-1}$. Every object has a \ssig~value measured from the old population that is consistent with that of the K star average.

Figure \ref{mgdisp} compares the measurements from the \ion{Mg}{1b} region, analogous to Fig.~\ref{baldisp} for the Balmer region. The left panel shows the comparison between the \ssig~measured for the young population and that of the old population. However, unlike in the case of the Balmer region, we find no correlation between these two measurements. The right panel of the figure shows that the \ssig~value of the old population is consistent with those values measured in the Ca T region, but the correlation has an average offset of $23\pm{35}$ km s$^{-1}$. The \ion{Mg}{1b} region is dominated by weak stellar features that are primarily attributed to the older stellar population. The only strong line constraining the young population is the H$\beta$ absorption line. It seems likely that because the young population is not well constrained here, the \ssig~measurements of that component are not trustworthy.

The bottom panels of Fig.~\ref{mgdisp} show the comparison between the K+A templates and the population models for the \ion{Mg}{1b} region, analogous to Fig.~\ref{baldisp} for the Balmer region. We find that the \ssig~of the young population is again overestimated by a single A star compared to using a population model (fitted intercept and slope $32\pm15$ km s$^{-1}$ and $0.96\pm{0.10}$, respectively). This result is similar to that of the Balmer region. Furthermore, the measurements of \ssig~for the old population are very consistent between the K star template and the old population template (fitted intercept and slope $-16\pm13$ km s$^{-1}$ and $1.12\pm{0.09}$, respectively).

\begin{figure*}
\epsscale{1.0}
\begin{tabular}{cc}
\epsfig{file=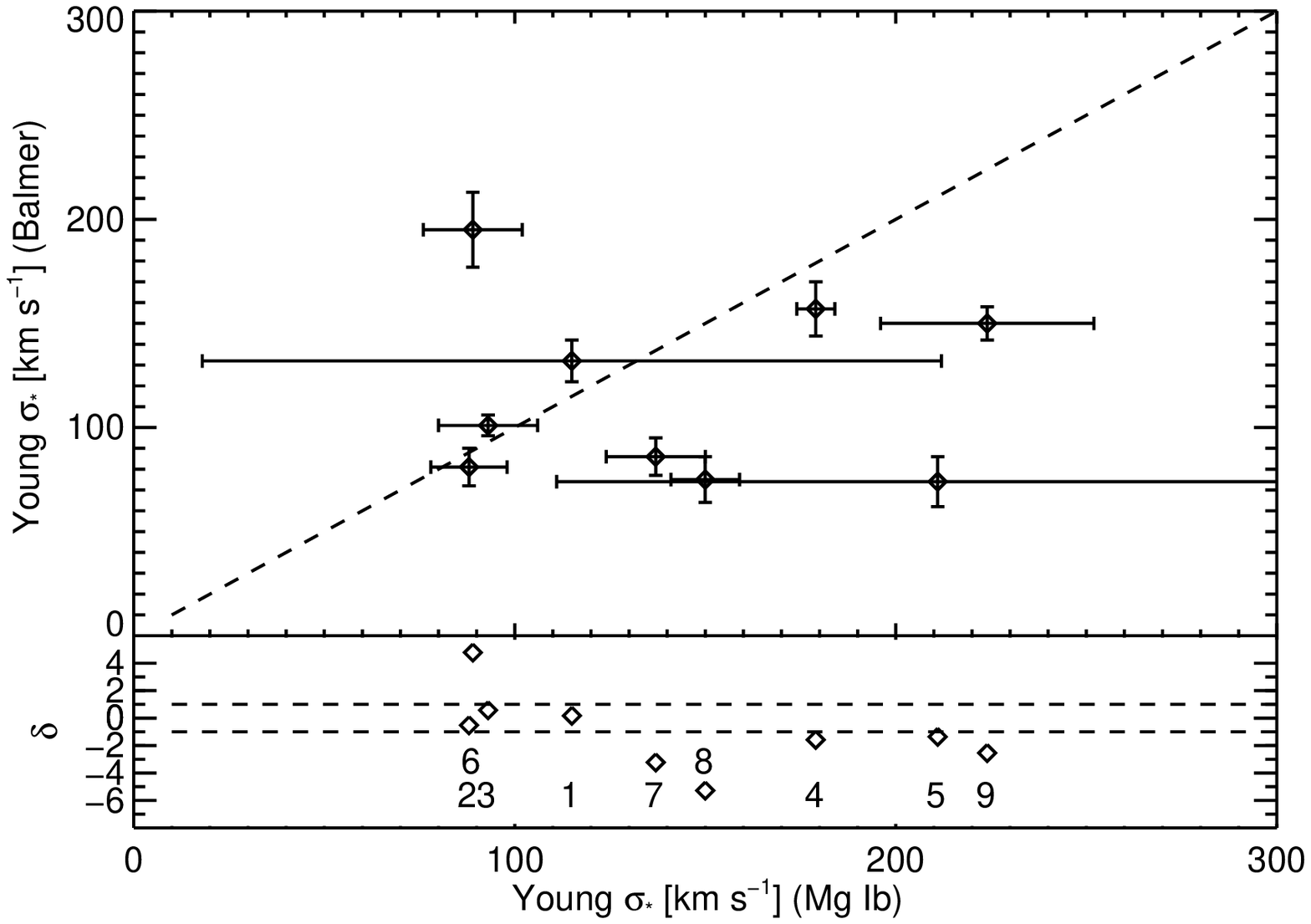, width=0.5\linewidth,clip= } & \epsfig{file=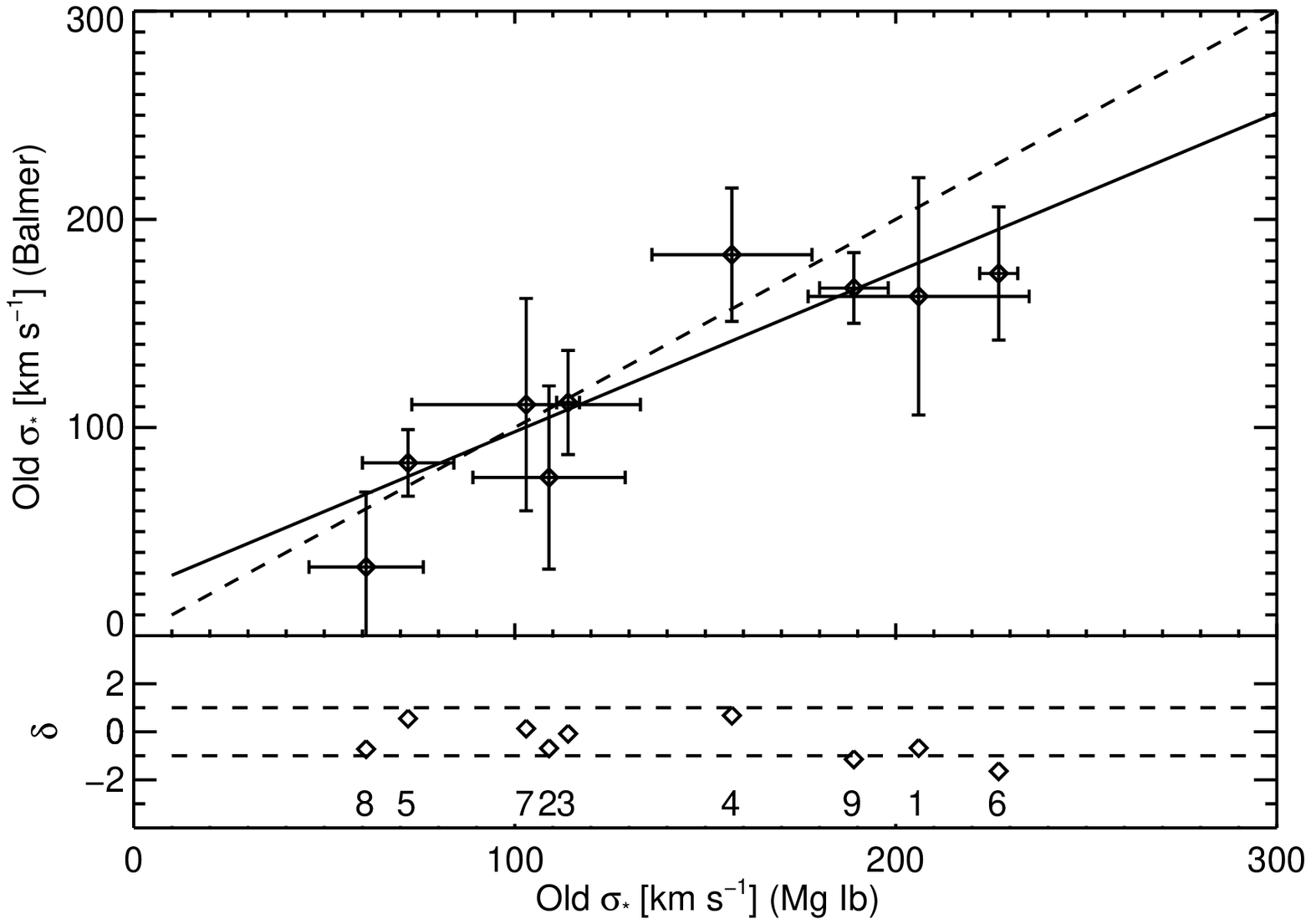, width=0.5\linewidth,clip= }\\
\end{tabular}
\caption{Velocity dispersion comparisons between measurements from the Balmer region and from the \ion{Mg}{1b} region. In each panel the dotted line has slope of one, and the solid line is a fit to the data. The fitted parameters are shown in Table \ref{fitparams}. No correlation is seen between the young population \ssig~measurement from the Balmer region with that of the \ion{Mg}{1b} region. However, the old population \ssig~measurements are correlated and consistent with a one-to-one relation. }
\label{balvmg}
\end{figure*}

\begin{table*}\centering
\caption{Linear fits to the \ssig~measurement comparisons \label{fitparams}}
\begin{tabular}{lccccc}
\hline
 Figure & intercept & slope & \chisq & \chisq/$\nu$ & q \\
\hline
8 - {\it Top Left} & $4\pm{28}$ & $0.90\pm{0.22}$ & 3.1 & 0.52 & 0.88 \\
8 - {\it Top Right} & $34\pm{35}$ & $0.95\pm{0.28}$ & 4.2 & 2.10 & 0.24 \\
8 - {\it Bottom Left} & $38\pm{16}$ & $0.91\pm{0.13}$ & 5.6 & 0.93 & 0.58 \\
8 - {\it Bottom Right} & $8\pm{13}$ & $1.01\pm{0.10}$ & 7.5 & 1.25 & 0.37 \\
9 - {\it Top Right} & $23\pm{35}$ & $0.96\pm{0.25}$ & 1.2 & 0.6 & 0.76 \\
9 - {\it Bottom Left} & $32\pm{15}$ & $0.96\pm{0.10}$ & 16.0 & 2.67 & 0.03 \\
9 - {\it Bottom Right} & $-16\pm{13}$ & $1.12\pm {0.09}$ & 6.1 & 1.02 & 0.52 \\
10 - {\it Right} & $21\pm{25}$ & $0.77\pm{0.17}$ & 3.3 & 0.55 & 0.86 \\
\hline
\end{tabular}
\end{table*}

Figure \ref{balvmg} compares the \ssig~measurements of the young population as measured in the Balmer and \ion{Mg}{1b} regions in the left panel, and those of the old population in the right panel. In the left panel there is no apparent correlation between the dispersions of the young populations as measured in the two wavelength regions. As stated above, some objects show H$\alpha$ emission, which raises the possibility that the remaining Balmer lines are also subject to emission-line filling. This would have the largest effect on the H$\beta$ line. Object SDSS $2041-0513$ (No. 5) exhibits the most filling and also shows the largest difference between \ssig~values measured from the Balmer and \ion{Mg}{1b} regions. This casts further doubt on the young population \ssig~value measured in the \ion{Mg}{1b} region. The Balmer region contains many more absorption lines, which are less affected by emission-line filling, to constrain the fit for the young population, so it is more trustworthy. We do see a correlation between the measurements of the older population. The slope of the linear fit is $0.77\pm{0.17}$, the lowest value for any of our fits. We note that the fit is not robustly determined and including more objects may significantly alter the fit. 

All of the relations discussed in this section depend on the adopted uncertainties of the individual \ssig~values. Increasing or decreasing the uncertainty on individual measurements can significantly change the values of the slopes or intercepts of the determined relations. However, we are only looking to see if measurements of various populations are consistent with one another. While this requires reasonably determined errors, which we believe we have adopted, it does not require such a robust and statistically significant sample that the intercepts and slopes of the linear fits are determined to high precision. In general, as summarized in Table \ref{fitparams}, we find slopes that are consistent with a one-to-one relationship between the measurement methods. The majority of offsets are small compared to their determined errors, but two are significant at the 2$\sigma$ level: the A star measurements compared to the young population template in both the Balmer and \ion{Mg}{1b} regions. These offsets indicate that using a single A star overestimates the velocity dispersion compared to a population template. 

In summary, the above figures show that the young and old stellar populations have similar kinematics and also highlight some potential issues with measuring velocity dispersions. For example, using only one Balmer line to constrain the young population is unreliable, and using a single A star to model the young population can bias results. We will discuss the results further in section \ref{conclude}.

\subsection{Faber-Jackson Relations}\label{FJsection}

The Faber-Jackson relation is well established for dynamically stable passively evolving galaxies \citep{FJ76,Desroches07,Nigoche10}. The relation is a projection of the Fundamental Plane and is a result of the virial theorem and the mass-to-luminosity ratio of stars and galaxies. We can use the Faber-Jackson relation to test if the objects are in dynamical equilibrium, or if the kinematics are peculiar compared to spheroids that are in dynamical equilibrium. 

We used the SDSS magnitudes along with our measured redshifts, K-corrections \citep{Chilingarian10}, and passive evolution corrections \citep{Leitherer14, Leitherer99} to derive absolute magnitudes in the $g$ and $r$ bands. The K-corrections account for differences in measured luminosities due to redshift and spectral color. We used the online calculator\footnote{http://kcor.sai.msu.ru/} described by \citet{Chilingarian10} and input the SDSS magnitudes and colors to derive K-corrections. When necessary we converted Cousins V and B band values to SDSS g and r bands using the conversions by \citet{Jester05}.

Passive evolution corrections account for the fading of stellar populations as they evolve between the observed redshift and $z=0$. The redshifts of our sample are quite low ($z < 0.1$) with the exception of one object, which has $z=0.2$. Previous authors have used corrections of 1.0 mag in Cousins R band for samples of similar redshifts \citep{Norton01, Pracy09}. To calculate passive evolution corrections for our sample, we created luminosity evolution tables of stellar populations using Starburst 99\footnote{http://http://www.stsci.edu/science/starburst99/docs/default.htm} \citep{Leitherer99} with the default model parameters of the simulation, but include the most recent evolution tracks as described in \citet{Leitherer14}. We calculated the evolution corrections for each population (young and old) individually based on the redshift of the galaxy and the age corresponding to the best fit templates. The old population was always simulated with a 10.0 Gyr population, while the young population was best fit in 7 galaxies by our 286 Myr template and the remaining 2 by the 202 Myr template. 

The passive evolution correction for the total galaxy is a combination of the corrections for the individual populations and depends on their flux ratio. We measured the relative flux contributions for the individual components from our fitted models and calculated the evolution correction for the total galaxy.  Corrections for the young population ranged from 0.66 to 1.76 mag in the $g$ band and from 0.27 to 1.27 mag in the $r$ band, with mean values 0.92 and 0.54 mag, respectively. 

The uncertainty in the evolved luminosity is dominated by the passive evolution correction, because the correction depends on the determined age of the young population and the fractional contribution to the overall luminosity. We calculated what the correction on the total magnitude would be if the young population were younger than we estimated and if it were older than we estimated by an amount equal to the age difference between stellar population templates that we tested (\ie~453 Myr - 286 Myr = 167 Myr). This produces an asymmetric error bar, because simple stellar populations evolve more at younger ages than they do at older ages. Additionally, after an initial fading, simple stellar populations brighten at ages ~1 Gyr before fading again. The passive evolution corrections for our sample fall right around this bump due to the redshifts of the objects. The corrections on the total luminosity are less affected by the bump in brightness, because total luminosity has significant contributions from the older population ($\sim30\%$ in $g$ and $\sim46\%$ in $r$).

The $g$ band photometry could range from 0.07 mag fainter to 0.01 magnitudes brighter due to uncertainties of the fitted stellar population ages and passive evolution corrections; the $r$ band photometry could range from 0.16 magnitudes fainter to 0.03 magnitudes brighter. These values are larger for the young population alone ($g$ and $r$ bands $\sim0.6$ fainter to $0.13$ mag brighter), but the uncertainty on the total magnitude is proportionately weighted by the contribution from the older population ($g$ and $r$ bands $\pm0.015$). These uncertainties are larger than the photometric errors from the SDSS, which are $< 0.01$ mag.

\begin{figure*}
\epsscale{1.0}
\begin{tabular}{cc}
\epsfig{file=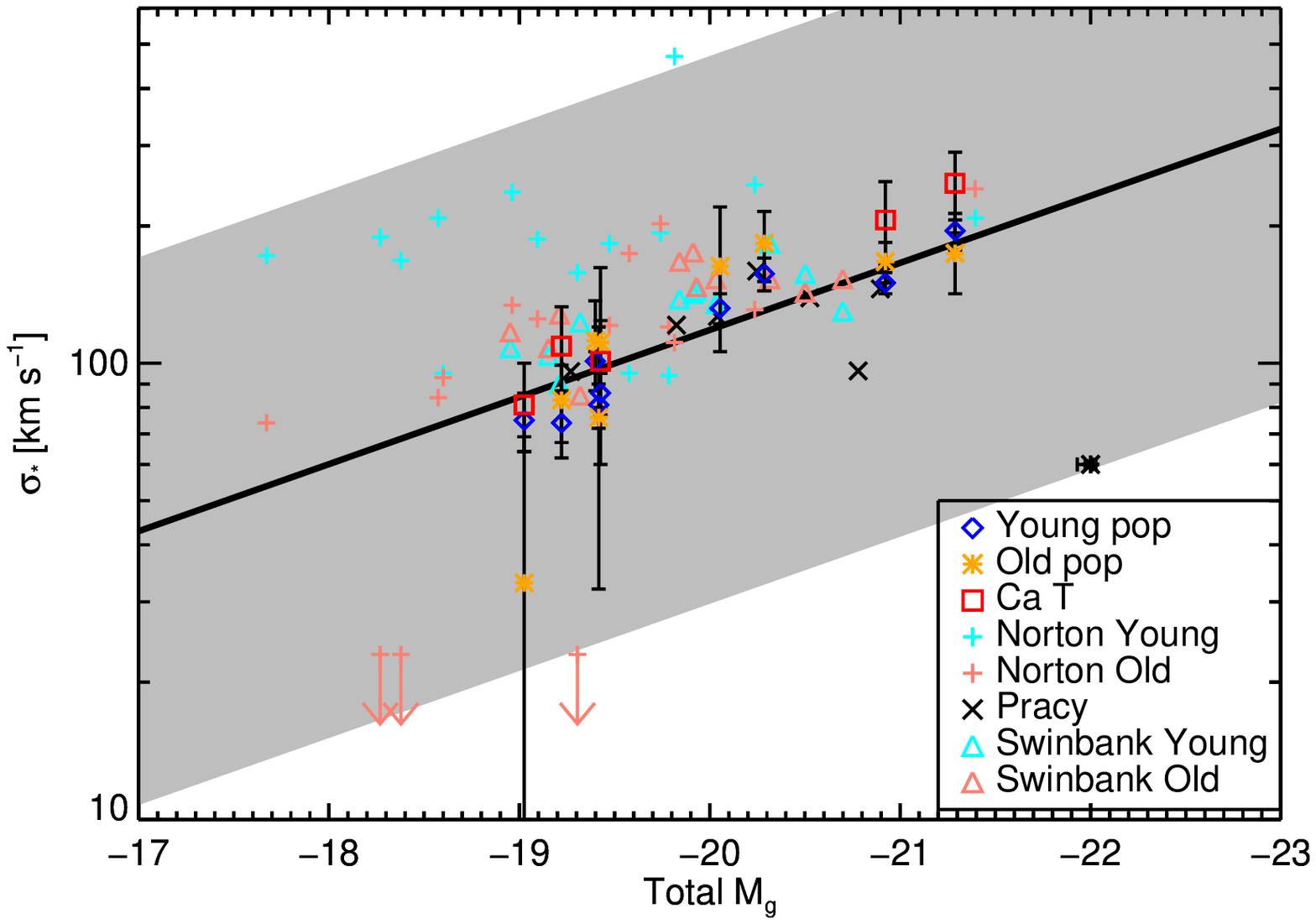, width=0.5\linewidth,clip= } & \epsfig{file=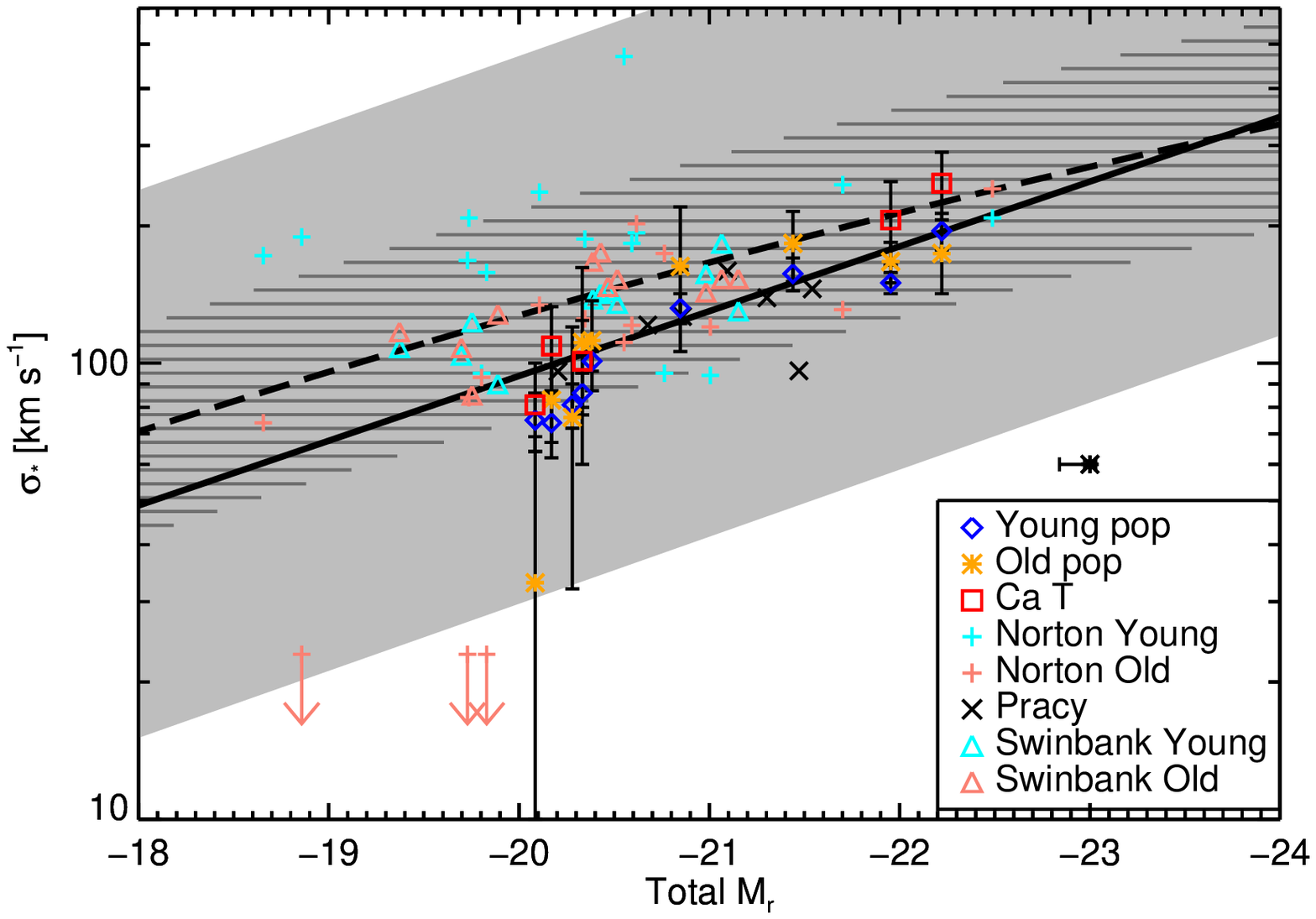, width=0.5\linewidth,clip= }\\
\end{tabular}
\caption[Faber-Jackson relation]{The Faber-Jackson relation for post-starburst galaxies. The velocity dispersions shown are as measured from the Balmer region for the young (blue diamonds) and old (orange asterisk) populations. The Ca T measurements (red squares) are also shown. The black asterisk point shows a representative error bar for the magnitudes determined by the uncertainties in passive evolution corrections (see text). {\it Left}: The velocity dispersions compared to the total $g$ band absolute magnitude. {\it Right}: The velocity dispersions compared to the total $r$ band absolute magnitude. In both panels, the solid line and filled region are a fit of the Faber-Jackson relation and scatter from \citet{Nigoche10}, while the dashed line and hashed region are a second-order fit and scatter of the relation by \citet{Desroches07}. Comparison objects from \citet{Norton01}, \citet{Pracy09}, and \citet{Swinbank12} are also shown. }
\label{FJtotal}
\end{figure*}

In Fig.~\ref{FJtotal} we show the Faber-Jackson relation for the post-starburst galaxies. We plot the SDSS $g$ band absolute magnitude in the left panel, and the $r$ band magnitude in the right. Measurements of \ssig~are from the Balmer region, which we consider to produce the most reliable measure of both the young and old population velocity dispersions. Blue diamonds represent the young population measurements, and orange asterisks represent the old population measurements. We also plot the \ssig~values measured from the Ca T region with red squares. The solid line represents the overall fit to the Faber-Jackson relation by \citet{Nigoche10}, which has a scatter of 0.6 dex. Our measurements of the individual populations are consistent with the relation. The SDSS $r$ band provides a slightly different picture (right panel Fig.~\ref{FJtotal}).

In the $r$ band, we compare our objects to the relation determined by \citet{Desroches07} (dashed line), who measure the Faber-Jackson relation in the SDSS $r$ band and perform a second-order fit that characterizes the relation's dependence on luminosity. We use the uncertainty in the fitted constant, the dominant uncertainty, to define the hashed region in the right panel of Fig.~\ref{FJtotal}. We find that the majority of our measurements from the Balmer region are consistent with the relation by Desroches \etal. Although, some of our objects fall on the lower cusp of the Desroches \etal~scatter region. For reference, the \citet{Nigoche10} relation for the $r$ band is also plotted as a solid line, and our objects show a similar behavior relative to that fit as they do in the $g$ band. 

For comparison purposes, we have included objects from \citet{Norton01}, \citet{Pracy09}, and \citet{Swinbank12} in Fig.~\ref{FJtotal}. These previous researchers use different photometric bands than we do here. To be consistent, we have used SDSS photometry for the objects in the previous studies where available, and treated the objects in the same way that we treat our sample. In the case of the \citet{Norton01} sample, SDSS photometry was unavailable. In this case, we used the photometry from the Las Campanas Redshift Survey \citep{Schectman96} and the APM survey \citep{Maddox90} along with photometric conversions provided by \citet{Blanton01} to calculate the appropriate SDSS magnitudes. For all of the comparison objects, we adopt a passive evolution correction of 1.0 mag, which is the value that the previous authors used. Velocity dispersion measurements were either published in the previous papers or provided by private communication with the authors. 

\citet{Norton01} plot their post-starburst galaxies on the M$_{\rm R} - $\ssig~plane and find that the young population of their sample did not closely follow the established Faber-Jackson relation. Their objects can be seen in Fig.~\ref{FJtotal} denoted as plus symbols. The young population shows significant scatter and appears offset to larger \ssig~values. The old populations, on the other hand have general agreement with the established relations. In comparing our results with those from \citet{Norton01}, it is important to note the differences in method of measuring \ssig. Norton \etal~fit two small spectral regions from $4100-4500$ \AA~and $4800-5250$ \AA. The first region contains H$\delta$,  H$\gamma$, and the G-band, while the second contains the H$\beta$ and \ion{Mg}{1b} lines. Furthermore, they only use the following spectral types: A2, A3, G4, G8, K0, and K3, including no F-type stars. We have compared the differences of using an average A star with a population template and found that using a single spectral type can bias the measurements of \ssig~for the young population. Another complicating factor is the passive evolution correction. While we perform an analysis of the young and old contributions to the fluxes, Norton \etal~use a value of 1.0 mag for the entire sample. A more detailed analysis of the relative fluxes of their sample may alter the objects' locations on the diagram. 

Similarly, \citet{Pracy09} show their 10 local post-starburst galaxies on the M$_{\rm R} - $\ssig~plane and find that they are largely consistent with the defined relation. The sample from Pracy \etal~is plotted on Fig.~\ref{FJtotal} as black X marks, and fall on the linear fit by \citet{Nigoche10} in both the $g$ and $r$-bands. They do not differentiate between the velocity dispersions of the young and old components. \citet{Swinbank12} plot their sample of post-starburst galaxies on the M$_{\rm K} - $\ssig~plane using \ssig~measured from SDSS spectra. While many of their objects are consistent with the K-band relation, several show an offset toward lower \ssig~values. Using available SDSS photometry of their sample, we have included them in Fig.~\ref{FJtotal} as triangles. In our plots, the objects are certainly within the 1$\sigma$ scatter of the linear relation by \citet{Nigoche10}, and fall directly on the second-order fit by \citet{Desroches07}. We note that we used a passive evolution correction of 1.0 mag for the \citet{Pracy09} and \citet{Swinbank12} objects, so the same caveat applies to their samples as to the sample studied by \citet{Norton01}. Overall, the comparison objects show good agreement with the previous fits, with the possible exception of the young component measurements by Norton \etal.

\begin{figure*}
\epsscale{1.0}
\begin{tabular}{cc}
\epsfig{file=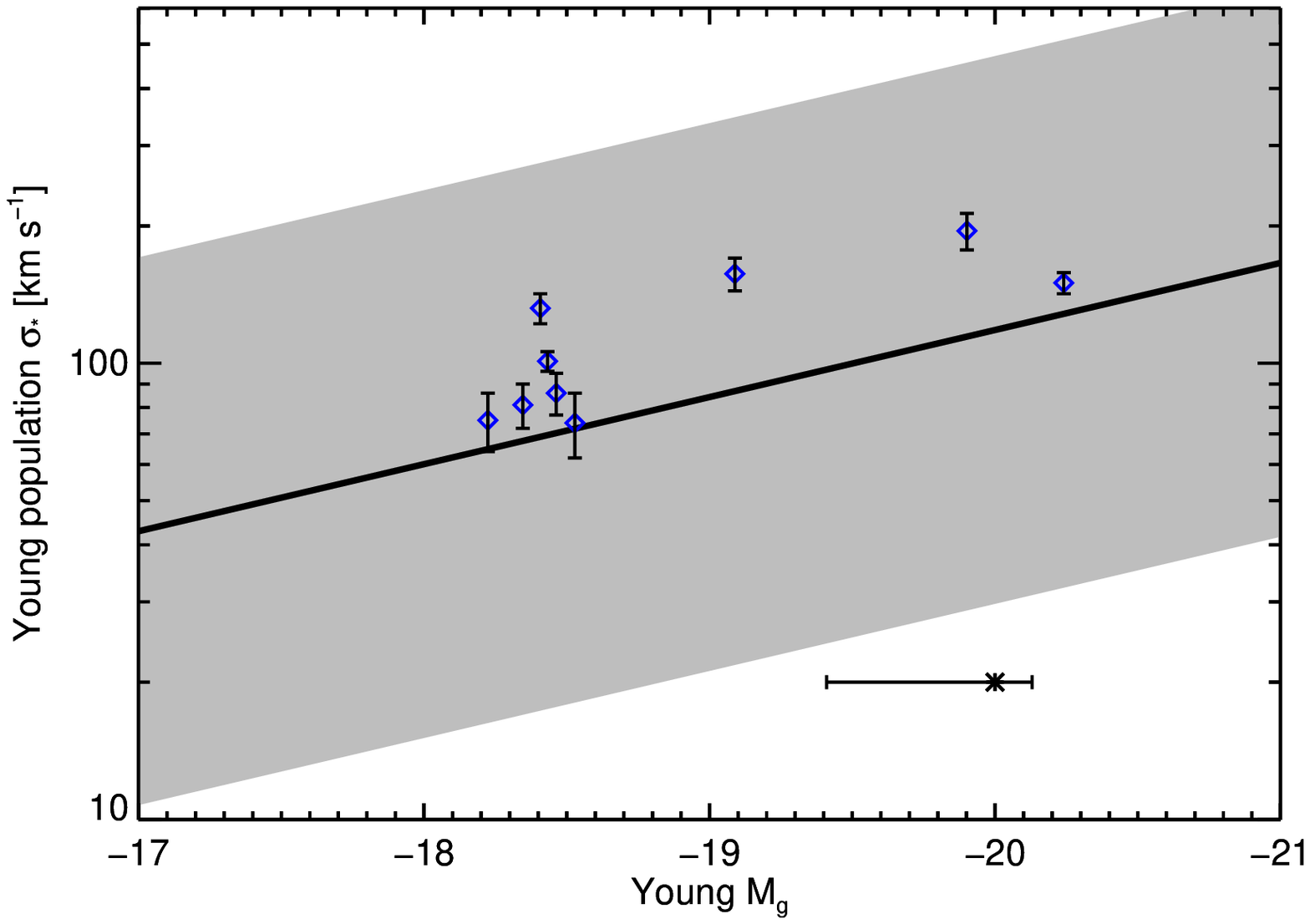, width=0.5\linewidth,clip= } & \epsfig{file=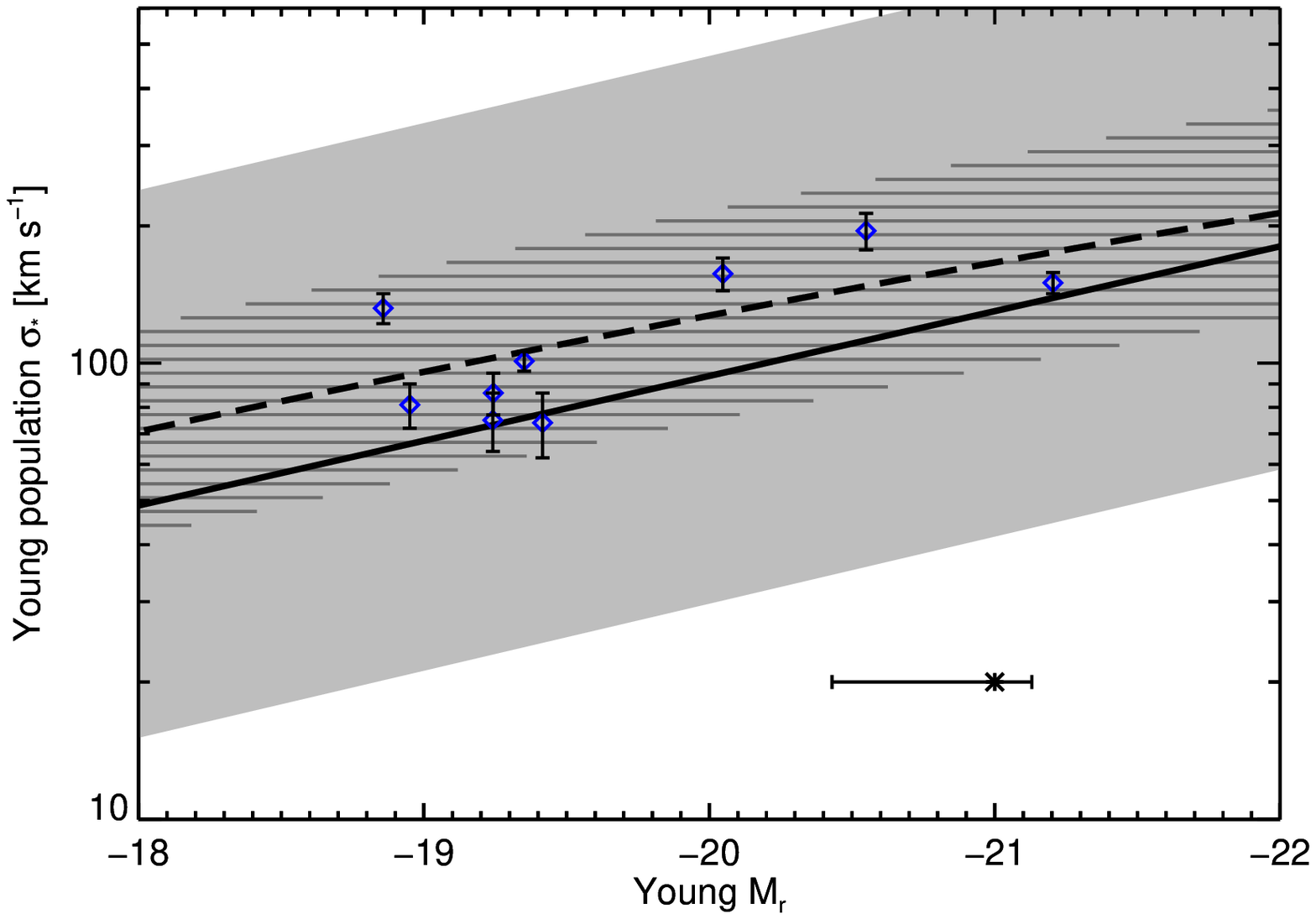, width=0.5\linewidth,clip= }\\
\epsfig{file=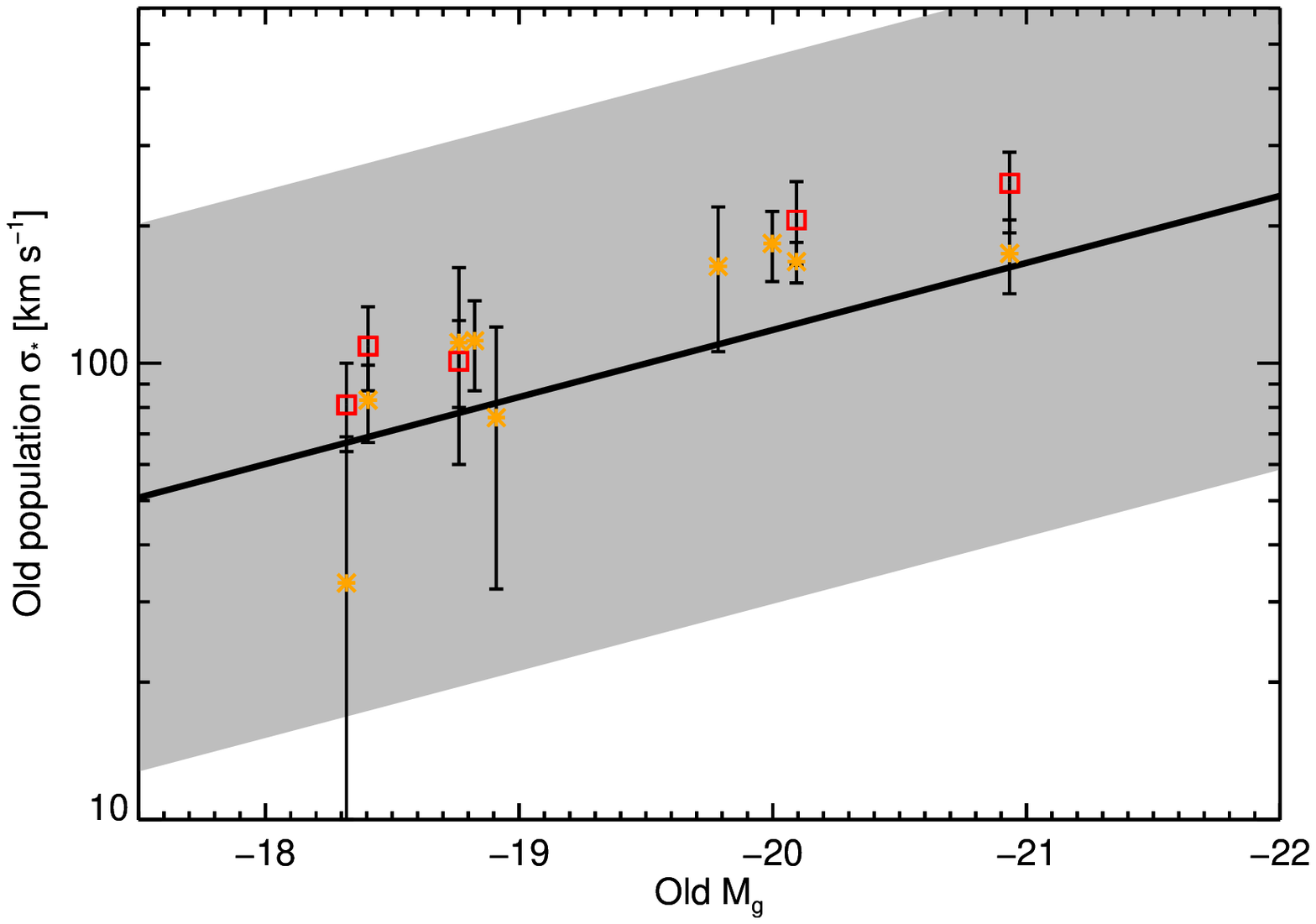, width=0.5\linewidth,clip= } & \epsfig{file=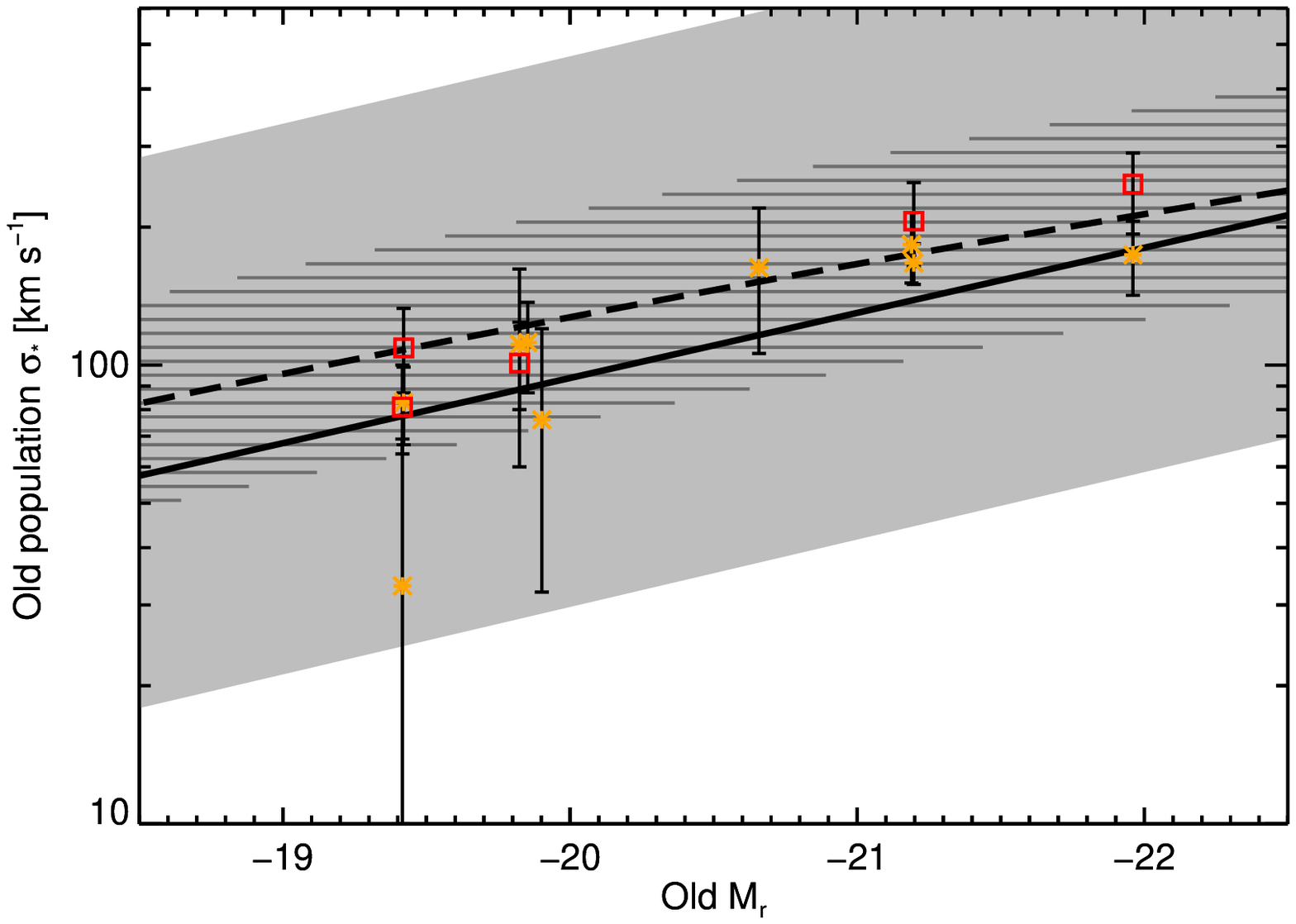, width=0.5\linewidth,clip= }\\
\end{tabular}
\caption[Faber-Jackson relation young and old components]{The objects plotted on the Faber-Jackson relation by their respective contributions to the luminosity. {\it Top}: Measurements of the young population \ssig~from the Balmer region compared to that population's contribution to the host luminosity in the $g$ and $r$ bands. The black asterisk point shows a representative error bar for the magnitudes determined by the uncertainties in passive evolution corrections (see text). {\it Bottom}: Measurements of the old population \ssig~from the Balmer region compared to that population's contribution to the host luminosity in the $g$ and $r$ bands. The photometric error bars on the old population are smaller than the mark size. The solid line and filled region are a fit of the Faber-Jackson relation and scatter from \citet{Nigoche10}, while the dashed line and hashed region are a second-order fit and scatter of the relation by \citet{Desroches07}. Even when considering contributions to luminosity only from the individual populations and the respective \ssig, the objects are still consistent with the Faber-Jackson relation.}
\label{FJcomponents}
\end{figure*}

We further investigated the dependency of \ssig~on the galaxy magnitudes by considering the individual contribution of each stellar population to the overall luminosity. We obtained initial photometric measurements of our total models and the individual stellar populations by integrating the product of the $g$ and $r$ band transmission curves with our normalized model spectra. We then calculated photometric scaling factors between our normalized photometry and the absolute $g$ and $r$ band luminosities (uncorrected for passive evolution) from SDSS. We then applied that scale factor to photometry of the individual young and old component populations. Lastly, we applied individual passive evolution corrections to the populations in both photometric bands. The results of this decomposition are the young and old components' individual contributions to the total galaxy magnitude as they would appear after passively evolving. Naturally, the individual component magnitudes are fainter than the total magnitude.

We plot the young component \ssig~values against the young component magnitudes in the top panels of Fig.~\ref{FJcomponents}. In the $g$ band, several of the objects appear above the \citet{Nigoche10} relation, but they still fall within the scatter. In the $r$ band, the measurements are still consistent with the \citet{Desroches07} quadratic fit. This result is largely similar to that of the total integrated light, which is perhaps not surprising as the young component contributes significantly to the total luminosity. We plot the older component magnitudes in the bottom panels of Fig.~\ref{FJcomponents} along with the older component \ssig~values and find similar results. In the $g$ band, the objects fall on or above the Faber-Jackson relation, but well within the scatter of the fit. In the $r$ band, the objects are still consistent with the fitted relation, with the exception of one point that falls below the relation by Desroches \etal. In summary, whether examining the stellar populations as a whole or the individual components, the post-starburst galaxies are consistent with the Faber-Jackson relations. As shown in Fig.~\ref{FJtotal}, previously studied objects are also largely consistent with the Faber-Jackson relation. This is an indication that they are not strongly perturbed away from a virial equilibrium.

\subsection{Fundamental Plane}

\begin{figure}
\plotone{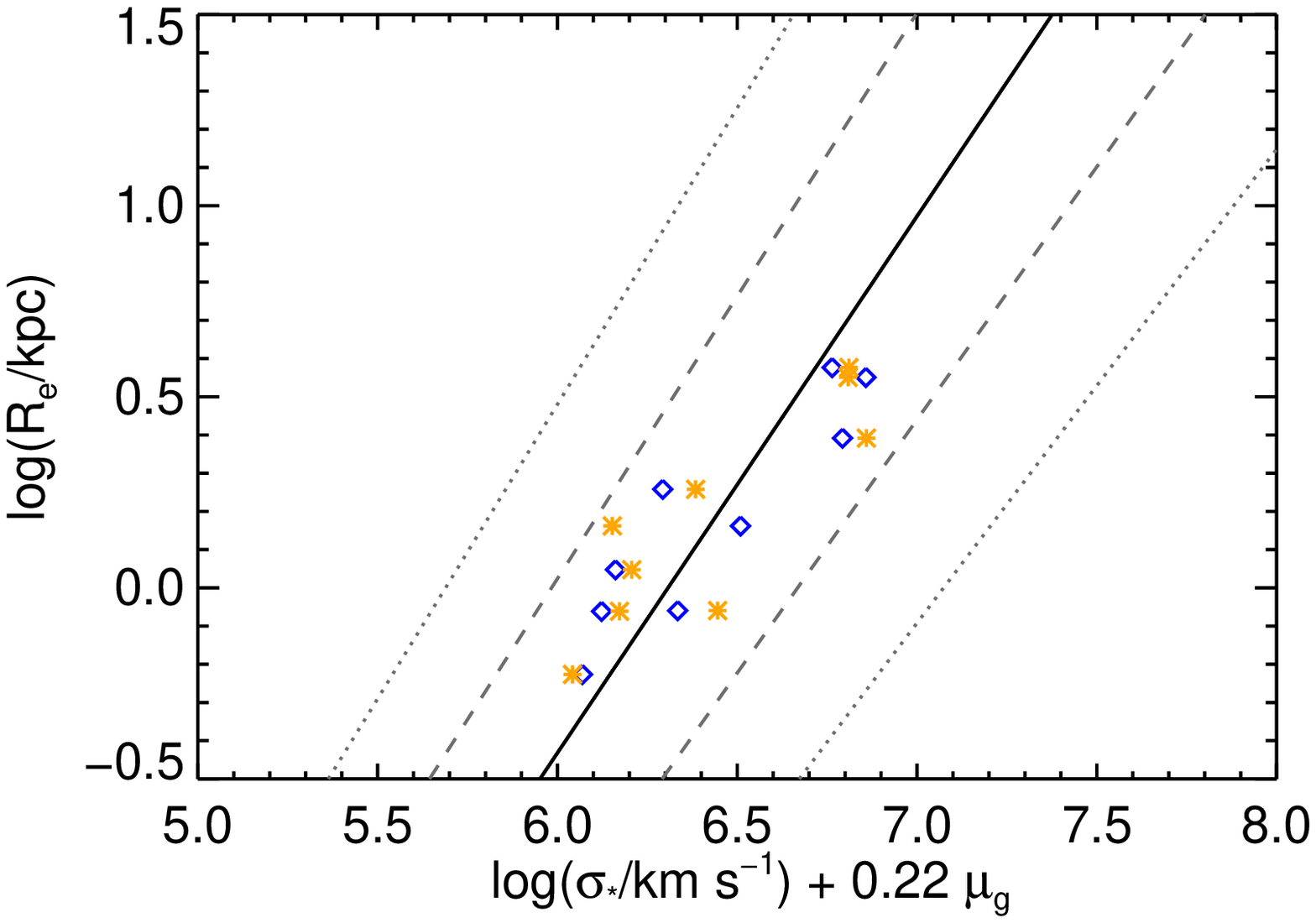}
\caption[Fundamental Plane]{The Fundamental Plane in the $g$ band. The solid line represents the fit from \citet{HB09}. The dashed and dotted lines encompass the rms and twice the rms of the scatter in the direction orthogonal to the fit, respectively. The blue diamonds represent the post-starburst galaxies of our sample using the young \ssig~measured from the Balmer region. The orange asterisks show the same objects, but use the \ssig~value of the older population. The \ssig~values have been corrected to \ssig/8 for purpose of comparison to \citet{HB09} only.} 
\label{fig:FP}
\end{figure}

In Fig.~\ref{fig:FP} we plot our post-starburst galaxies on the Fundamental Plane. For comparison we use the relation determined by \citet{HB09} from $\sim$50000 early-type galaxies from the SDSS. Their relation is in the form log(R$_e$) = a~log(\ssig/km s$^{-1}$) + b~$\mu_{\rm eff}$ + c. The scale constant c is determined from the average values of R$_e$, \ssig, and $\mu_{\rm eff}$. We use their fit for the SDSS $g$ band: ${\rm a} =1.4043$, ${\rm b} =0.3045$. Systematic errors dominate the uncertainties and yield $\delta{\rm a} =0.05$ and $\delta{\rm b} =0.02$. We plot the plane similarly to Hyde \& Bernardi, the only difference being that they have plotted log(\ssig) + 0.22($\mu$ - $<\mu>$) and we have plotted log(\ssig) + 0.22($\mu$).

We measured surface brightnesses of the post-starburst galaxies using the $g$ band DR8 SDSS images of our sample and the 2D modeling routine GALFIT \citep{Peng10}. In the modeling routine, we restricted our fits to S\'{e}rsic profiles, letting the index be a free parameter. The indices converged in the range 1.3 - 4, with all of the fits producing reasonable residuals. Bulges with n$\sim$2 are sometimes classified as pseudo-bulges. In our analysis we corrected the surface brightness for cosmological dimming effects, but note that this was a minor correction given the low-redshifts of our sample. We also applied the same passive evolution correction to the total $g$ band luminosity as we did in Section \ref{FJsection}.

Again, we adopt our measured values of \ssig~from the Balmer region for the plot. In Fig.~\ref{fig:FP}, the blue diamonds are plotted using the young stellar population \ssig, while the red squares reflect the older population dispersion. \citet{HB09} use an aperture correction to transform the \ssig~values of their sample to $\sigma_{e8}$. Since our aperture extraction was defined at R$_e$, not R$_{e}/8$, we have corrected our measurements found in Table \ref{vdisp} to $\sigma_{e8}$ using the prescription from \citet{JFK95} for comparison on Fig.~\ref{fig:FP} only. 

We can characterize the general uncertainty of the objects' locations on the Fundamental Plane as follows: The SDSS seeing and pixel scale affect the precision of the R$_e$ measurement. For one object (SDSS $0119+0107$, No. 1), the FWHM of the PSF corresponds to a physical scale of 2.23 kiloparsec, and log(FWHM) of $\sim$0.35. Since we perform a fit to the SDSS image, the uncertainty on the R$_e$ parameter is smaller than the FWHM of the PSF. Meanwhile, the uncertainty on \ssig~dominates over that of the surface brightness measurement. The range of \ssig~values corresponds to an uncertainty along the horizontal direction that ranges from $\sim$0.6 to $\sim$1.5. 

Given such error bars on the horizontal axis, we find that the post-starburst galaxies are fully consistent with the Fundamental Plane as derived by \citet{HB09}, and certainly within the 1$\sigma$ uncertainties of their fits to the relation. This is further evidence that the kinematics of the young population are not peculiar relative to their host. Our post-starburst galaxies are generally in dynamical equilibrium similar to evolved early-type galaxies.

\section{A Case Study of $2337-1058$}\label{case}

We serendipitously observed the galaxy SDSS $2337-1058$ (No. 9) with the slit oriented very close to the major axis of the galaxy (see Fig.~\ref{img}). This provides an opportunity to study the internal kinematics of the two populations and to determine if the observed ellipticity of the galaxy is due to rotational flattening or if the galaxy is largely pressure supported. From the SDSS imaging, it is clear that this object is actively undergoing a merger event. A companion galaxy can be seen just to the North of the main galaxy, and a tidal tail extends to the SW. Our slit was oriented roughly NE-SW along the major axis. The object lies at $z=0.0783$, providing an angular scale of 1.462 kpc arcsec$^{-1}$. 

We measured the ellipticity of the galaxy by fitting a S\'{e}rsic profile to the main galaxy in GALFIT using the SDSS $g$ band image. We used a star located on the same image frame as the PSF to be convolved with the S\'{e}rsic profile. The best-fit function left residuals extending between the tidal tail and the companion galaxy. We measured a S\'{e}rsic index of 3.1 and an axis ratio of $b/a = 0.43$, corresponding to an ellipticity of $\epsilon = 0.90$. 

To determine if this ellipticity can be produced by rotational flattening of the galaxy, we measured the line-of-sight velocity, $v_{\rm los}$, and velocity dispersion, \ssig, of eleven aperture extractions along the major axis of the galaxy. Each aperture extraction was offset from its neighboring aperture by the resolution limit of our seeing conditions, $0\farcs6$, corresponding to a physical separation of 0.877 kpc per aperture. We performed the same spectral fitting on the Balmer region as described in Section 3. We note that because of the size of the galaxy most of the eleven aperture extractions are within the measured effective radius of the galaxy, which was 3.76 kpc (2.57 arcsec). Only the outer two apertures fall outside the effective radius.

\begin{figure}
\plotone{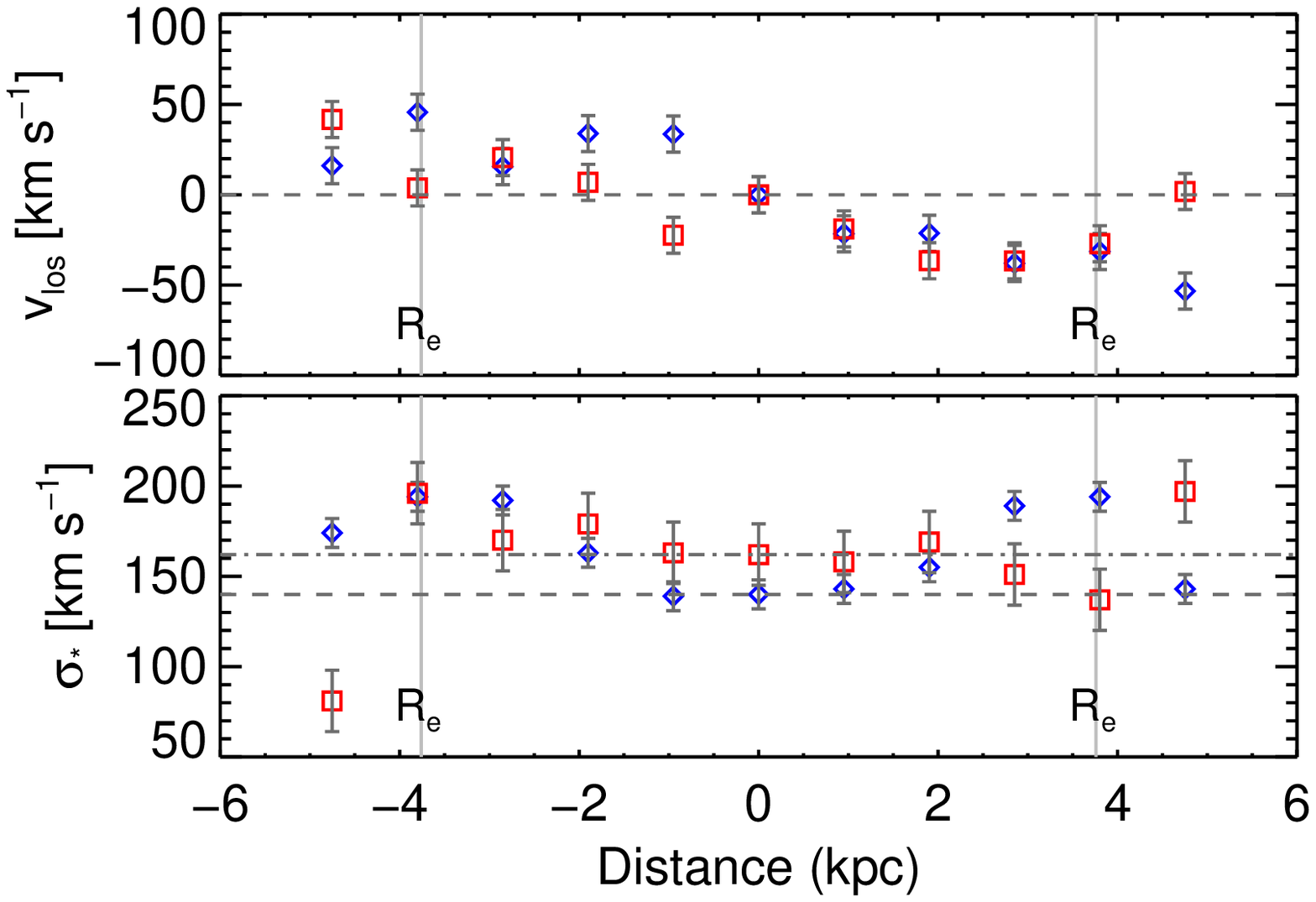}
\caption[Internal kinematics of $2337-1058$]{Kinematics of SDSS $2337-1058$ (No. 9). The young population kinematics are plotted as blue diamonds, and the old population as red squares. {\it Top}: The line-of-sight velocity is plotted relative to the value from the central aperture. {\it Bottom}: The velocity dispersion as measured in each aperture. Each aperture is separated by the seeing limit 0\farcs6. The physical scale is 1.462 kpc arcsec$^{-1}$. In both panels the vertical solid lines denote the location of one effective radius. The horizontal dashed lines indicate the values of v$_{\rm los}$ and \ssig~at the central aperture. }
\label{aps2337}
\end{figure}

Table \ref{2337} shows the measured $v_{\rm los}$ and \ssig~values for both the young and old components of the galaxy using the Balmer region. Figure \ref{aps2337} shows the values plotted as a function of position within the galaxy. From the figure it is clear that some rotation is present in the galaxy. Line-of-sight velocities for both populations, but moreso for the young, are positive along the NE direction of the slit and negative along the SW direction of the slit relative to the central aperture. The \ssig~values of the young population are slightly depressed in the innermost region of the galaxy. This could indicate the presence of a rotating disk composed primarily of the young population at the nucleus of the galaxy, which would be consistent with models from \citet{Bekki05}. Furthermore, the young component \ssig~is smaller than that of the older component in the innermost apertures. Bekki \etal~also find this offset in both major and unequal-mass merger models, although the magnitude of the effect is much greater in their simulations.

\begin{table*}\centering
\caption{Kinematics of $2337-1058$ \label{2337}}
\begin{tabular}{cccccc}
\hline
Aperture & $v_{\rm los}^{y}$ & $v_{\rm los}^{o}$ & $\sigma^{y}$ & $\sigma^{o}$ & Young template age \\
\hline
01 & 16.1 & 41.7 & 174 & 81 & 718 \\
02 & 45.7 & 3.8 & 194 & 196 & 453 \\
03 & 15.6 & 20.6 & 192 & 170 & 453 \\
04 & 33.9 & 6.9 & 163 & 179 & 286 \\
05 & 33.6 & -22.4 & 139 & 163 & 286 \\
06 & 0.0 & 0.0 & 140 & 162 & 286 \\
07 & -21.6 & -18.9 & 143 & 158 & 286 \\
08 & -21.3 & -36.5 & 155 & 169 & 286 \\
09 & -38.0 & -36.6 & 189 & 151 & 286 \\
10 & -31.4 & -27.1 & 194 & 137 & 453 \\
11 & -53.3 & 1.8 & 143 & 197 & 718 \\
\hline
\end{tabular}
\tablecomments{The stellar population systemic velocities and velocity dispersions as measured from eleven aperture extractions along the major axis of $2337-1058$. Aperture 06 is the central aperture and the other apertures extend in the NE - SW directions. All velocities are presented in km s$^{-1}$, and the template ages are in Myr. }
\end{table*}

To determine if the observed ellipticity is due to rotational flattening, as might be the case for a galaxy that collapsed from a single primordial gas cloud, we calculate the anisotropy parameter: 

\begin{equation}
(v/\sigma)^* = \frac{(v/\bar{\sigma})}{\sqrt{(\epsilon/(1-\epsilon))}}
\end{equation}

\noindent
where $v$ is the maximum observed line-of-sight velocity (centroid of absorption lines), $\bar{\sigma}$ is the average of the velocity dispersions within the effective radius and $\epsilon$ is the ellipticity of the galaxy \citep{BenderNieto90, Kormendy82}. If $(v/\sigma)^* \sim 1$, then ordered rotational motion dominates and the observed flattening is consistent with a disk structure. If, on the other hand, $(v/\sigma)^* << 1$, then the random motion of the stars ($\sigma$) is sufficiently high to rule out such a rotational flattening, and the object has an anisotropic \ssig~tensor. In the case of SDSS $2337-1058$, $(v/\sigma)^* = 0.091$ and $0.084$ for the young and old components, respectively, indicating that random motions of stars are dominant over ordered motions in this system.

Another test of the merger-driven origin hypothesis is to examine the spatial distribution of the young population within the galaxy. Several authors have previously investigated Balmer line gradients with various results. \citet{Pracy13} note that merger models indicate the young population should get funneled to the central $\sim 1$ kpc of the galaxy \citep[\eg][]{Bekki05,Snyder11,Stickley14}. This should be revealed in a peak of H$\delta$ equivalent width in the innermost regions of the galaxy. Our observations are seeing limited, which projects to just over one-half kiloparsec per resolution element (0$\farcs$6) at the redshift of SDSS $2337-1058$ ($1.462$ kpc arcsec$^{-1}$). This allows us to just resolve the innermost region and test whether the starburst is driven to the center of the galaxy by a merger event.

\begin{figure}
\plotone{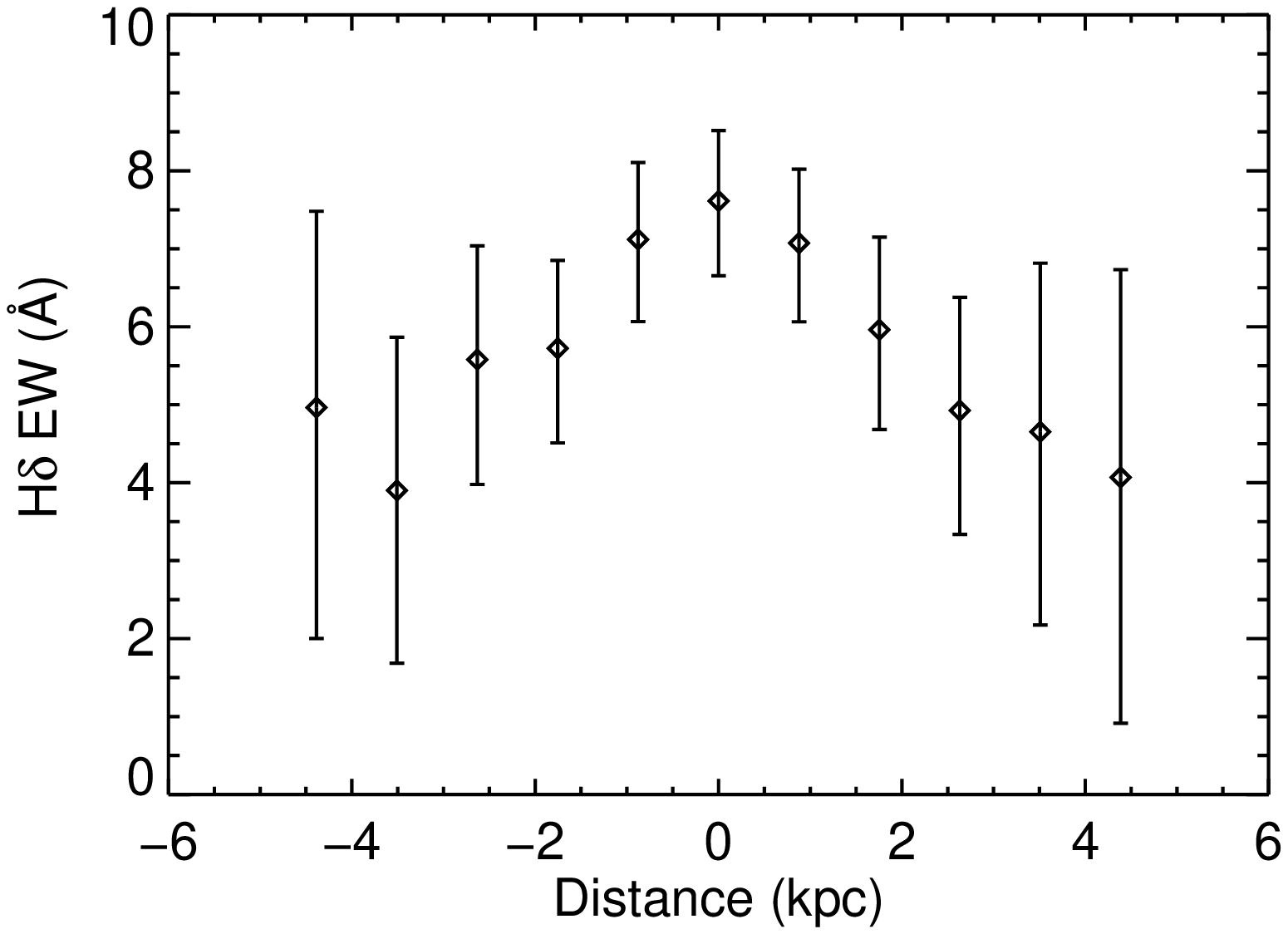}
\caption[H$\delta$ Gradient]{The H$\delta$ equivalent widths as a function of radius from the galaxy center for SDSS $2337-1058$ (No. 9). As in Fig.~\ref{aps2337}, each aperture is separated by the seeing limit of 0\farcs6, and the scale is 1.462 kpc arcsec$^{-1}$. The gradient of the equivalent widths indicates the young population is centrally located within this galaxy.}
\label{fig:Hdelta}
\end{figure}

In Fig.~\ref{fig:Hdelta} we show the H$\delta$ equivalent widths (EW) as a function of radius from the center of the galaxy SDSS $2337-1058$. We measured the EW using the Lick index definition from \citet{WO97}. The line was integrated over wavelengths $4083.50-4122.25$ \AA, the blue continuum was measured from the average flux between $4041.60-4079.75$ \AA, and the red continuum was measured from the average flux between $4128.50-4161.00$ \AA. We estimated errors by determining the flux dispersion across the continuum regions. We then adjusted the continuum level with the dispersions to determine maximum and minimum values of the EWs. 

Fig.~\ref{fig:Hdelta} clearly shows a gradient with respect to radius in the H$\delta$ EWs. This is consistent with results from \citet{Pracy13}, who find a slight increase in H$\delta$ EWs in the innermost regions of local post-starbursts using data from an IFU. The observed gradient in SDSS $2337-1058$ suggests that the young population is centrally concentrated within the inner kiloparsec of the galaxy, a result that might naturally arise from a merger scenario.

\section{Summary and Discussion}\label{conclude}

We have measured velocity dispersions (\ssig) of different stellar populations within a sample of nine post-starburst galaxies selected from \citet{Goto07}. The stellar kinematics of post-starburst galaxies can be directly observed by fitting population templates to observed spectra. Furthermore, with sufficiently high signal-to-noise, it is possible to distinguish between young and old populations. While spectra of young populations are dominated by Balmer transitions in the optical, older populations have many less prominent features, of which the Ca H+K lines and G band are strongest. 

Below we summarize our results with a series of recommendations for fitting \ssig~in post-starburst galaxies:

\begin{enumerate}

\item Measurements of the young population are best constrained in the Balmer region. Including the majority of the Balmer series absorption lines in the fit provides a much stronger constraint on the population kinematics than using only a single spectral feature (\eg~H$\beta$ near the \ion{Mg}{1b} region). The older population can be fit simultaneously with the younger population, even in the Balmer region where the luminosity is dominated by young stars. In this region there still exist distinct spectral features and a multitude of weaker spectral features that arise from the older population. As long as the spectra are sufficiently high signal-to-noise, these features can constrain the older population parameters reasonably well.

\item It is important to fit the galaxy spectrum with a young population model rather than a single A star. Line ratios vary between early and late A-type stars, and F stars contribute significantly to the Balmer series. Thus, using a single A star can cause a systematic bias in the measured dispersion. In both the Balmer region and the \ion{Mg}{1b} region, the older population is modeled equally well using either a population model or a single representative K star average. This is likely simply due to the fact that older populations are dominated by the later type stars, which all have similar features.

\item When measured with an aperture that spans the effective radius, R$_e$, the kinematics of the young population are consistent with those of the old population as measured in the Balmer region of the spectrum. This indicates the two populations are dominated by the same dynamics and potential well. The kinematics of the older population are more uncertain than those of the young population due to template mis-match, because a degeneracy develops between the old population and the ``aged'' young population templates where later-type stars become more prevalent. It is reasonable to fit both populations together to probe the overall galaxy potential. 

\item The \ssig~values measured from the Ca T region are more closely aligned with old population \ssig~from the \ion{Mg}{1b} region than the Balmer region. Interestingly Fig.~\ref{baldisp} implies that the \ssig~values of both the young and old populations as measured from the Balmer region under-estimate the value as measured from the Ca T region. One should be aware of the ``\ssig~discrepancy" when planning observational programs. The magnitude of the discrepancy is still uncertain, but can be on the order of $\sim$30 km s$^{-1}$.

\end{enumerate}

The ``\ssig~discrepancy" has been studied by several authors. In their simulations of major mergers, \citet{Stickley14} recover the discrepancy and posit a rough physical interpretation: Stars formed in the early stages of the merger between passes of galaxies may have lower \ssig, because they tend to be formed in low-dispersion tidal features that later coalesce with the galaxies as the merger progresses. However, the central starburst during the final stages of the merger happens in a dynamically cold disk near the nucleus, which results in stars with lower dispersion compared to the host. The stars formed near the center and at the end of the merger do not have the benefit of the galaxy back-and-forth passages. As a result their velocity dispersion remains depressed with respect to the dispersion of the host. This physical interpretation may explain the observed discrepancy between the \ssig~as measured between the Balmer and Ca T regions. While the Balmer region is dominated by young and intermediate aged stars (early-type), the Ca T region is dominated by later-type (likely older) stars. Thus, the Ca T region may simply probe the pre-merger population or the stars formed in the earliest merger stages that have mixed the most, and have the highest \ssig.

The discrepancy has been previously found in (ultra-) luminous infrared galaxies \citep[LIRGs and ULIRGs;][]{Rothberg10,Rothberg13} but not for objects of lower infrared luminosities \citep[\eg][]{Vanderbeke11,Kang13}. \citet{Rothberg13} observe the discrepancy in ULIRGs by measuring \ssig~from the Ca T region and the 1.6 and 2.3 $\micron$ CO infrared stellar absorption lines. In that case, the \ssig~determined from the Ca T was larger than the \ssig~determined by the CO lines. \citet{Rothberg10} propose that the discrepancy arises because the CO emission penetrates a layer of dust to observe a young central stellar disk in the nucleus of the LIRG/ULIRG. If this is the case, their results would be consistent with ours and with the interpretation posited by \citet{Stickley14}. That is, younger stellar populations tend to have lower \ssig~values than the older populations. 

We performed a case study of the object SDSS $2337-1058$, examining the major axis kinematics and the distribution of the young population using the H$\delta$ EW. This object is undergoing a merger, and both the young and old populations show some rotation. However, both the populations' velocity dispersions are sufficiently high to indicate that the object is pressure dominated. The curve of the young population \ssig~with respect to position within the galaxy is consistent with results from simulations of major mergers found by \citet{Bekki05}, showing a slight depression of \ssig~in the inner $\sim$2 kpc. This is an indicator the young population may be slightly disk-y, which would be consistent with the implications of the ``\ssig~discrepancy''. The extent of any diskiness would be better confirmed with data from IFU observations. Furthermore, we find that the young population is extended within the central few kiloparsecs of the galaxy. Despite being extended, it is certainly concentrated within the inner 1 kpc, further consistent with merger models that drive star-forming gas inward. 

The post-starburst galaxies measured here appear to be consistent with both the Faber-Jackson relation and the Fundamental Plane, indicating that they are not so tidally disrupted that their stellar motions are non-virial, nor are any potential disk components dominant over the pressure supported components. Thus, in post-starburst galaxies the Balmer region provides a measurement of stellar kinematics for both the young and old populations that is a reasonable probe of the overall galaxy potential. The fact that these systems are consistent with the above relations could be a clue to their origin. We would not expect systems that are in the early stages of a major merger to be highly consistent with the Fundamental Plane. In addition, the morphologies of these galaxies, while they occasionally show some tidal features or companions, are not highly disturbed like seen in the early stages of major mergers (\eg~multiple nuclei). Thus, these systems are likely in either a late-stage merger, or a more minor merger that does not disrupt the kinematics very much. This result is consistent with many of the aforementioned merger simulations that show significant star formation in the late stages of the mergers. 

Studying the dynamics of young populations of post-starburst galaxies can reveal formation mechanisms and provide clues to galaxy evolution. We note that recent studies with integral field units (IFU) have uncovered early type galaxies with distinct kinematical differences. New classifications of ``slow" and ``fast" rotators based on the $\lambda_{R}$ parameter \citep{Emsellem07} are now possible, which may provide further insight into the dynamical differences of young and old populations \citep[\eg][]{Pracy13,Swinbank12,Pracy09}. These studies have shown that most post-starbursts are fast rotators, which is dissimilar from large elliptical galaxies, leading the authors to posit a scenario where gas-rich minor-mergers are fueling the starburst. Our results are consistent with that scenario. Post-starburst galaxies and those with active nuclei, while rare, could be an important puzzle piece uniting disk-y, star-forming galaxies with AGN and passively evolving ellipticals.

\acknowledgments

The authors would like to thank the anonymous referee for providing valuable suggestions that improved the paper. The authors also thank M. Pracy and A. M. Swinbank for sharing data from previous works, and the authors thank S. L. Cales for productive discussions. 

While conducting this research K.~Hiner was supported by the UCR Dissertation Year Fellowship and FONDECYT grant 3140154. This research was supported in part by NASA through a grant from the Space Telescope Science Institute (program number AR-12626).

The data presented herein were obtained at the W.M. Keck Observatory, which is operated as a scientific partnership among the California Institute of Technology, the University of California and the National Aeronautics and Space Administration. The Observatory was made possible by the generous financial support of the W.M. Keck Foundation.The authors wish to recognize and acknowledge the very significant cultural role and reverence that the summit of Mauna Kea has always had within the indigenous Hawaiian community.  We are most fortunate to have the opportunity to conduct observations from this mountain.

\newpage

\bibliographystyle{apj}
\bibliography{ref}

\end{document}